\documentclass[apj]{emulateapj}
\usepackage[colorlinks,linkcolor={blue},citecolor={blue},urlcolor={red}]{hyperref}
\usepackage{epsfig,graphicx,natbib,amsmath,amsfonts,amssymb}
\usepackage{color}
\usepackage{multirow}
\usepackage{booktabs}
\usepackage{threeparttablex} 
\usepackage{booktabs} 

\usepackage{longtable}
\usepackage{array} 

\newcommand{\re}{\Reff}
\newcommand{\msolar}{M$_\odot$}
\newcommand{\mstar}{M$_\ast$}
\newcommand{\lgmstar}{$\log_{10}$(M$_\ast/$\msolar)}
\newcommand{\dindex}{D$_n$(4000)}
\newcommand{\dindexin}{D$_n$(4000)$_{\rm cen}$}
\newcommand{\dindexout}{D$_n$(4000)$_{\rm 1.5Re}$}
\newcommand{\concen}{R$_{90}/R_{50}$}

\newcommand{\nuvr}{NUV$-r$}
\newcommand{\hd}{H$\delta$}
\newcommand{\hda}{\hd$_A$}
\newcommand{\ewhda}{EW(\hda)}
\newcommand{\ha}{H$\alpha$}
\newcommand{\hae}{\ha}
\newcommand{\ewhae}{EW(\hae)}
\newcommand{\lgewhae}{$\log_{10}$\ewhae}

\newcommand{\Reff}{{$R_{\rm e}$}}

\newcommand{\msun}{M$_{\odot}$}

\newcommand{\sersic}{S\'{e}rsic}

\newcommand{\myemail}{\email{ecwang16@ustc.edu.cn(EW), xkong@ustc.edu.cn(XK)}}

\shorttitle{Massive SF galaxies with Outside-in assembly mode}
\shortauthors{Wang et al.}

\graphicspath{{fig/}}

\begin{document}

\title{The Properties of Massive Star-forming galaxies with Outside-in Assembly Mode}
\author{
Enci Wang\altaffilmark{1,2},
Xu Kong\altaffilmark{1,2},
Huiyuan Wang\altaffilmark{1,2},
Lixin Wang\altaffilmark{3},
Lin Lin\altaffilmark{4},
Yulong Gao\altaffilmark{1,2},
Qing Liu\altaffilmark{1,2}
} \myemail

\altaffiltext{1}{CAS Key Laboratory for Research in Galaxies and Cosmology, Department of Astronomy, University of Science and Technology of China, Hefei 230026, China}
\altaffiltext{2}{School of Astronomy and Space Science, University of Science and Technology of China, Hefei 230026, China}
\altaffiltext{3}{Tsinghua Center of Astrophysics \& Department of Physics, Tsinghua University, Beijing 100084, China}
\altaffiltext{4}{Key Laboratory for Research in Galaxies and Cosmology, Shanghai Astronomical Observatory, 
Chinese Astronomical Society, 80 Nandan Road, Shanghai 200030, China}
\begin{abstract}  

Previous findings show that massive (\mstar$>10^{10}$\msun) star-forming (SF) galaxies usually have 
an ``inside-out'' stellar mass assembly mode.  In this paper, we have for the first time selected a sample
 of 77 massive SF galaxies with an ``outside-in'' assembly mode (called ``targeted sample'')
from the Mapping Nearby Galaxies at the Apache Point Observatory (MaNGA) survey.  
For comparison, two control samples are constructed from MaNGA sample matched in stellar mass: 
a sample of 154 normal SF galaxies and a sample of 62 quiescent galaxies. 
In contrast to normal SF galaxies, the targeted galaxies appear to be more smooth-like and bulge-dominated, 
and have smaller size, higher concentration, higher star formation rate and higher gas-phase metallicity
as a whole. However, they have larger size and lower concentration than quiescent galaxies. 
Unlike normal SF sample, the targeted sample exhibits a slightly positive gradient of  4000 \AA\ break and 
a pronounced negative gradient of H$\alpha$ equivalent width. 
Further more, their median surface mass density profile is between that of normal SF sample and quiescent sample,
indicating that the gas accretion of quiescent galaxies is not likely to be the main 
approach for the outside-in assembly mode.
Our results suggest that the targeted galaxies are likely in the 
transitional phase from normal SF galaxies to quiescent galaxies, with rapid on-going 
central stellar mass assembly (or bulge growth). 
We discuss several possible formation mechanisms for the outside-in mass assembly mode. 

\end{abstract}


\section{Introduction}
\label{sec:introduction}

 Thanks to large mount of imaging and spectroscopic survey, the knowledge of galaxies has drastically increased
  in past two decades.   Galaxies are mainly separated into two types according to its bimodality of color:  
   star-forming galaxies and quiescent galaxies \citep[e.g.][]{Baldry-04,Bell-04, Blanton-05}. 
  Star-forming galaxies typically have on-going star formation with disk-like morphology, while red quiescent 
  galaxies typically have no or very weak star formation activities with bulge-dominated morphology 
  \citep{Strateva-01, Kauffmann-03a, Baldry-04, Brinchmann-04, Li-06, Muzzin-13, Barro-17}. 
  However, the spatially resolved stellar mass assembly histories of galaxies are still poorly understood. 
  
  Galaxies appear to assemble their stellar mass mainly in two distinct modes: the ``inside-out'' and the 
  ``outside-in'' mode \citep{Pan-16, Perez-13, Li-15, Ibarra-Medel-16, Goddard-17}.
  In the inside-out assembly mode, star formation cessation occurs first in galactic center with respect to
  the outskirts, which may be due to active galactic nucleus feedback \citep[e.g.][]{Kauffmann-04, Bower-06, Fabian-12,
  Bournaud-14, Dekel-Burkert-14}, the ``morphological quenching'' \citep{Martig-09} or the environmental effects
   \citep[e.g.][]{Gunn-Gott-72, Moore-96, Conselice-03, Cox-06, Cheung-12, Smethurst-15, Wang-15}. 
  This inside-out stellar mass assembly mode has been found in massive disk/star-forming galaxies
   (\mstar$>10^{10}$\msun) both in low and high redshift \citep[e.g.][]{Munoz-Mateos-07, 
  Sanchez-Blazquez-07, Perez-13, Bezanson-09,vandeSande-13,Patel-13,Tacchella-15}. 
  While for less massive galaxies (\mstar$<10^{10}$\msun), the main galaxy assembly mode is very different, 
  which may be better interpreted in ``outside-in'' scenario 
  \citep{Bernard-07, Gallart-08, Zhang-12, Perez-13, Pan-15}. 
  
  Recently, several large integral field spectroscopy (IFS) surveys have been performed, such as ATLAS$^{\rm 3D}$
  \citep{Cappellari-11}, CALIFA \citep{Sanchez-12}, SAMI \citep{Bryant-15}, and MaNGA \citep{Bundy-15},
   which enables people to study the spatially resolved mass assembly histories of galaxies. 
  By applying the fossil record method on 105 CALIFA galaxies, \cite{Perez-13} have found 
  that the spatially resolved stellar mass assembly pattern of local galaxies depends on the global stellar mass.  
  Specifically, the massive galaxies grow their stellar mass inside-out, while less massive galaxies
  show a transition to outside-in growth. 
  In agreement with this, \cite{Pan-15} proposed that the typical galaxy assembly mode is transiting 
  from ``outside-in'' mode to ``inside-out'' mode from low to high stellar masses, 
  indicated by \nuvr\ color gradients of $\sim$10000 low-redshift galaxies.
 By analyzing 533 galaxies selected from MaNGA, \cite{Ibarra-Medel-16} have generated the global and radial 
 stellar mass growth histories with fossil record method and confirmed that the massive SF
  galaxies, on average, show a pronounced inside-out mass assembly mode. 
  They also found that the less massive galaxies show diverse mass assembly histories, 
  with periods of outside-in and inside-out modes.
\cite{GonzalezDelgado-15, GonzalezDelgado-16} generated the maps of mean stellar age 
and metallicity for 416 galaxies from CALIFA, and found that the negative 
gradients of light-weighted stellar age are consistent with inside-out growth of galaxies, 
especially for Sb-Sbc galaxies.
These findings agree well with the negative stellar age gradients for massive SF galaxies
 \citep[e.g.][]{Wang-11, Lin-13, Li-15, Dale-16, Goddard-17}.

In this paper, we for the first time select a small but significant fraction ($\sim$14\%) 
of massive star-forming galaxies (\nuvr$<$5) with recent outside-in mass assembly 
mode from the MaNGA released data \citep{-16}.
These galaxies are selected to have higher 4000 \AA\ break at the 1.5 times effective radius (\re) than 
at the galaxy center. The 4000 \AA\ break is commonly used to indicate the recent star formation 
histories within 2 Gyr \citep{Bruzual-Charlot-03, Kauffmann-03a}.
For comparisons, we build two control samples matched in stellar mass.   
  
This paper is organized as follows.  
 In Section \ref{sec:data}, we describe the data we used and sample selection. 
 We present our results in Section \ref{sec:results}, including the global properties, the resolved star formation histories, 
 and the environments of the targeted sample and two control samples.  
 In Section \ref{sec:discussion}, we discuss the possible formation mechanisms of targeted galaxies, and the implication of star formation    
 quenching based on our results. We summarize our results in Section \ref{sec:summary}. 
Throughout this paper, all the distance-dependent parameters are computed 
with $\Omega_m=0.3$, $\Omega_\Lambda=0.7$ and $\rm H_0=70\ km\ s^{-1}Mpc^{-1}$.

\section{Data}
\label{sec:data}

\subsection{MaNGA data overview}
\label{subsec:data_overview}

Our main sample is the MaNGA released data, including 1390 galaxies, which enables us to select a sample of 
the massive SF galaxies with outside-in assembly mode.
Here we briefly introduce the MaNGA data.  A full description of both observations 
and data reductions are presented in \cite{Bundy-15}.

The MaNGA project is an on-going large integral-field spectroscopic survey, aimed at mapping
10000 nearby galaxies with redshift from 0.01 to 0.14. The MaNGA instrument covers a wavelength range from 
3600 to 10300 \AA\ at a typical resolution R$\sim$2000. The absolute flux calibration is better than 5\% 
for more than 89\% of MaNGA wavelength range \citep{Yan-16}. 
The raw data were reduced using the MaNGA Data Reduction Pipeline, which was
developed by \cite{Law-15}. 

The MaNGA released sample is mainly composed by the ``Primary'' ($\sim$2/3) and the 
``Secondary'' ($\sim$1/3) samples \citep{-16}, which are defined by two radial coverage goals, with the former 
one of 1.5\re, and the later one of 2.5\re.
The released datacubes have a spaxel size of 0.5 arcsec, and the typical effective spatial resolution
 can be described by a Gaussian with an FWHM$\sim$2.5 arcsec.

\subsection{Spectral fitting}
\label{subsec:spectral_fitting}

The spectral fitting methods and parametric measurements are described in \cite{Li-15}. 
Here we briefly summarize the fitting procedure and the measurements of referred parameters. 
To fit the continuum, we adopt a chi-squared minimization fitting method developed by \cite{Li-05}, 
which can carefully mask the emission-line regions by iteration. 
The templates we use are galactic eigenspectra generated by principal component analysis method. 
This fitting code is efficient and stable, even for spectra with low signal to noise ratio (SNR). 
However, this code is not able to fit the line of sight velocity automatically. 
Thus we use the line of sight velocity measured by the public code, STARLIGHT \citep{CidFernandes-05},  
as input for each spectrum. The templates used in STARLIGHT fitting are 45 single stellar populations from
\cite{Bruzual-Charlot-03} with a \cite{Chabrier-03} initial mass function (IMF). 
The intrinsic stellar extinction law is from \cite{Cardelli-Clayton-Mathis-89} 
in both fittings. For each galaxy, we perform the spectral fitting to spaxels separately without binning scheme. 
Spectra with the continuum SNR$<$5 at 5500 \AA\ are not included in analysis in this work. 

We use the 4000 \AA\ break (\dindex), H$\delta$ absorption (\ewhda) and H$\alpha$ emission (\ewhae) 
equivalent width to trace the recent star formation histories \citep{Bruzual-Charlot-03,Kauffmann-03a, Li-15}, 
since these parameters are more model independent in contrast to mean stellar age and star formation rate (SFR). 
 \dindex\ and \hda\ index are measured based on the best fitting spectra, while the \ewhae\ is measured 
 base on the stellar component-subtracted spectrum by fitting a Gaussian profile to H$\alpha$ emission lines.
 A full description of spectral fitting and parametric measurements were presented in \cite{Li-15}. 

\subsection{Sample selection}
\label{subsec:sample_selection}

\begin{figure*}
  \begin{center}
    \epsfig{figure=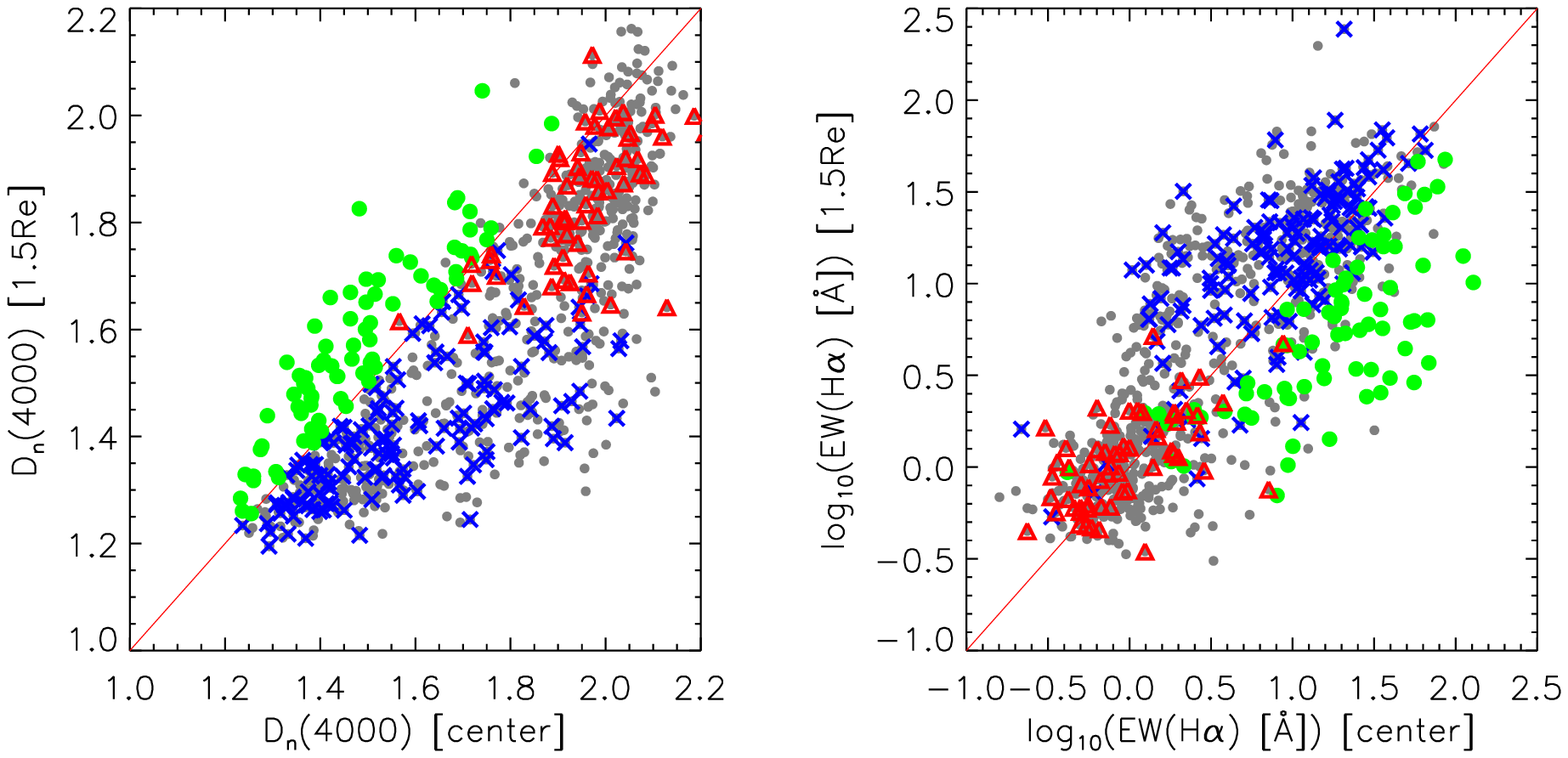,clip=true,width=0.6\textwidth}
    \epsfig{figure=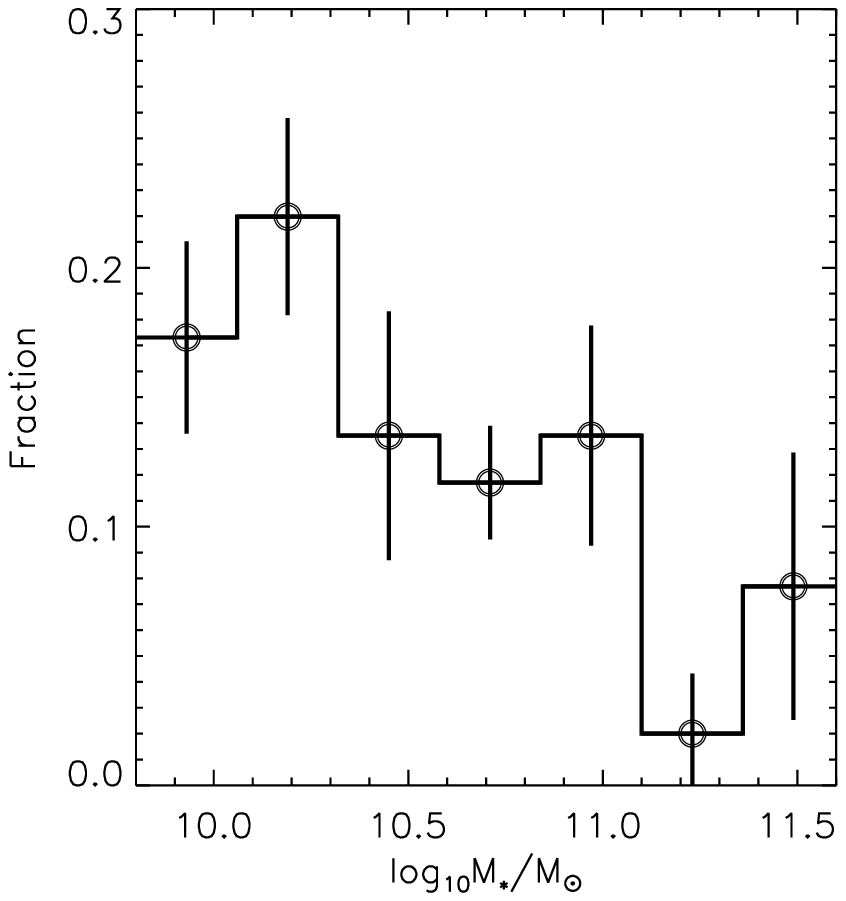,clip=true,width=0.3\textwidth}
  \end{center}
  \caption{Sample selection. Left panel: The comparison between 4000 \AA\ break at galactic center and at 1.5\re\ radial bin. Middle panel: The comparison between \lgewhae\ at galactic center and at 1.5\re\ radial bin. 
In each panel, the green points, blue crosses and red triangles represent the targeted galaxies, BC galaxies and RC 
galaxies, respectively. The red solid line is one-to-one line. 
For comparison, we plot all MaNGA galaxies as gray circles in the left two panels.
Right panel:  The fraction of the targeted galaxies in SF galaxies (\nuvr$<$5) as a function of stellar mass.
 The errors are estimated by bootstrap method. 
 }
  \label{fig:sample_selection}
\end{figure*}

In previous works, people usually separated galaxies according to their global properties, such as 
star formation rate (SFR), stellar mass and color. 
The IFS surveys show that different regions of galaxies can have very different star formation properties,
suggesting that only using global parameters is not enough.  
Thanks to the IFS survey, we are able to for the first time select the galaxies with outside-in
mass assembly mode and study their global and resolved properties. 

We use 4000 \AA\ break to select the targeted galaxies. The 4000 \AA\ break
is defined as the ratio of flux in red continuum to that in blue continuum at 4000 \AA\ wavelength 
\citep{Balogh-99, Brinchmann-04}. \dindex\ has been widely used to trace the mean stellar age of stellar 
populations younger than 1-2 Gyr.  We note that the mass assembly histories can be diverse in different 
periods \citep{Ibarra-Medel-16}, and the mass assembly mode referred in this work is recent
 assembly mode ($<$2 Gyr). 

 The targeted sample was selected from the SDSS-IV MaNGA released sample \citep{-16}, 
 which contains 1390 galaxies with 
 0.01$<$z$<$0.14.  Since we are just interested in massive galaxies with outside-in assembly mode, 
 the sample was required (1) to have \lgmstar$>$9.8, (2) to have larger 4000 \AA\ break 
 in 1.5\re\ radial bin (\dindexout) than in the center (\dindexin),
  (3) to be SF galaxies or partially SF galaxies (\nuvr$<$5).
  That we use  \mstar$>10^{9.8}$\msun\ rather than \mstar$>10^{10}$\msun, and \nuvr$<$5 
  rather than \nuvr$<$4, is to select more galaxies. This is useful for measuring the clustering 
  properties of our sample in Section \ref{subsec:environments}.  
 In addition, we have excluded mergers and heavily disturbed galaxies by visually
 inspection.  In total, 77 galaxies were selected according to the above criteria, which accounts for
 the proportion of 14\% of all galaxies with \lgmstar$>$9.8 and \nuvr$<$5. 
 
 For comparison, we have built two control samples closely matched in stellar mass with 
 targeted sample, but not in outside-in assembly mode.  
 The control samples were randomly selected from MaNGA released sample with several criteria. 
 We call them blue control (BC) sample and red control (RC) sample. The BC sample was selected 
 to have \dindexin$>$\dindexout\ and \nuvr$<$5, while the RC sample was selected 
 to have \nuvr$>$5. At last, 154 BC galaxies and 62 RC galaxies were selected. 

Since we do not limit the galaxy inclination in the sample selection, 12 out of 77 targeted galaxies
 have the inclination angle greater than 70 degree (based on the minor-to-major axis ratio from NSA).
  One may worry that the measurements of \dindex\ 
 are heavily affected by the dust extinction for these highly inclined galaxies, 
 which may pollute the identification of targeted galaxies.
  We have checked \dindex\ and \lgewhae\ profiles of these
  highly inclined galaxies, and find that they exhibit positive \dindex\ gradients, and prominently negative
   \lgewhae\ gradients in general. The measurements of \ewhae\ are not sensitive to the dust extinction, 
 which confirms that these highly inclined galaxies are in outside-in assembly mode. In addition, we have 
reproduced our main results with excluding these highly inclined galaxies, and find that the basic results are
 not changed. Thus, we keep these highly inclined galaxies in this paper.

The left two panels of Figure \ref{fig:sample_selection} show the targeted sample (green dots), BC sample (blue crosses)
 and RC sample (red triangles) on the diagrams of \dindexin\ vs. \dindexout\ and \lgewhae$_{\rm cen}$ vs.
  \lgewhae$_{\rm 1.5Re}$. 
For comparison, we plot all MaNGA galaxies with \mstar$>10^{9.8}$\msun\ as gray circles 
in these two panels. The right panel shows the fraction of our targeted galaxies with respect to massive 
SF galaxies (\nuvr$<$5 and \mstar$>10^{9.8}$\msun) as a function of stellar mass. 
The errors are estimated by bootstrap method.  The fraction of targeted galaxies
 has a peak at \mstar$\sim10^{10.2}$\msun, and drops heavily at \mstar$>10^{11.1}$\msun. 

As shown in Figure \ref{fig:sample_selection}, these three samples are well separated 
in the diagrams of \dindexin-\dindexout\ and \lgewhae$_{\rm cen}$-\lgewhae$_{\rm 1.5Re}$. 
 Most galaxies have higher 4000 \AA\ break in their centers than in 1.5\re\ radial bins, which is consistent
 with the previous findings of inside-out assembly mode for 
 massive galaxies \citep{Pan-15, Perez-13, Goddard-17}. 
 However, we do see a small fraction of massive SF galaxies with \dindexin$<$\dindexout, 
 shown in green dots. 
As expected, they have larger \lgewhae\ in their centers than in 1.5\re\ radial bins, indicating 
the dramatic on-going star formation activities in their centers. 
 In contrast, the BC galaxies have larger 4000 \AA\ break and smaller \ewhae\ in their 
 centers than in 1.5\re\ radial bins, representing normal SF population.
 The RC galaxies show no or very weak star formation activities, representing quiescent population. 
 They have larger 4000 \AA\ break and smaller \ewhae\ both in their centers and 
 1.5\re\ radial bins than targeted and BC galaxies.

\begin{figure*}
  \begin{center}
    \epsfig{figure=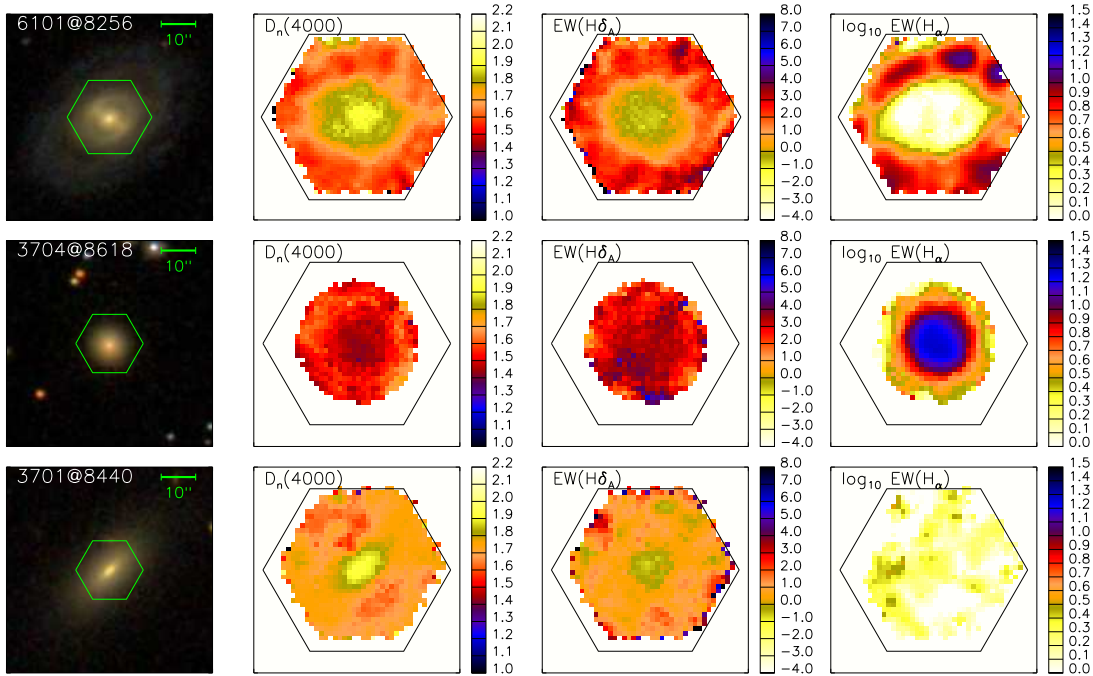,clip=true,width=0.85\textwidth}
  \end{center}
  \caption{Examples of 2-dimensional maps of SDSS images and diagnostic parameters. The top row shows 
  an example of a galaxy in BC sample with inside-out assembly mode. The middle row shows 
  an example of a galaxy in targeted sample with outside-in assembly mode. The bottom row shows 
  an example of a galaxy in RC sample. 
 In each row, the columns show the SDSS color image, 4000 \AA\ break, \hda\ index and \lgewhae\ maps from left to right. The hexagons on SDSS color image and diagnostic parameter maps represent the covered
 area by IFU bundles for each galaxy.}
  \label{fig:examples}
\end{figure*}

 Examples of a BC galaxy (top row), a targeted galaxy (middle row) and a RC galaxy 
(bottom row) are shown in  Figure \ref{fig:examples}.  
In each row, the columns represent the SDSS color image, the 4000 \AA\ 
break, \hda\ index and \lgewhae\ maps of the shown galaxy. The hexagon in each panel displays the
covered area by IFS bundles.  
\hda\ is Lick/IDS index of H$\delta$ absorption line \citep{Worthey-Ottaviani-97} and
\lgewhae\ is the logarithm of H$\alpha$ emission line equivalent width. These three parameters are good
indicators of recent star formation histories \citep{Bruzual-Charlot-03, Kauffmann-03a, Li-15}. 
\hda\ index traces the star formation that occurred 0.1-1 Gyr ago,
and \ewhae\ traces the strength of very recent (within 50 Myr) or on-going star formation activities.

As shown in Figure \ref{fig:examples}, the 2-dimensional star formation properties of the targeted 
galaxy (3704@8618, the galaxy located on plate 8618 with the bundle ID of 3704) 
 and the BC galaxy (6101@8256) are significantly different. The BC galaxy has
 higher \dindex, lower \hda\ index 
and \lgewhae\ in its center than its outer regions. The star formation activities have already been shut
 down in its center, where a bulge and a bar can be seen. 
 According to its \lgewhae\ map, on-going star formation activities can be seen at its outer regions.  
In contrast, the targeted galaxy (3704@8618) have lower \dindex, higher \hda\ index and \lgewhae\ in its 
center than its outer regions. It has a smooth-like morphology without spiral arms seen from the SDSS color
image. It has strong star formation activities in the center, while very weak star formation activities 
are shown in its outer regions. 
The RC galaxy (3701@8440) is a quiescent galaxy with no prominent H$\alpha$ emission lines. 

\section{Results}
\label{sec:results}

\subsection{Global Properties}
\label{subsec:global_properties}

\begin{figure*}
  \begin{center}
    \epsfig{figure=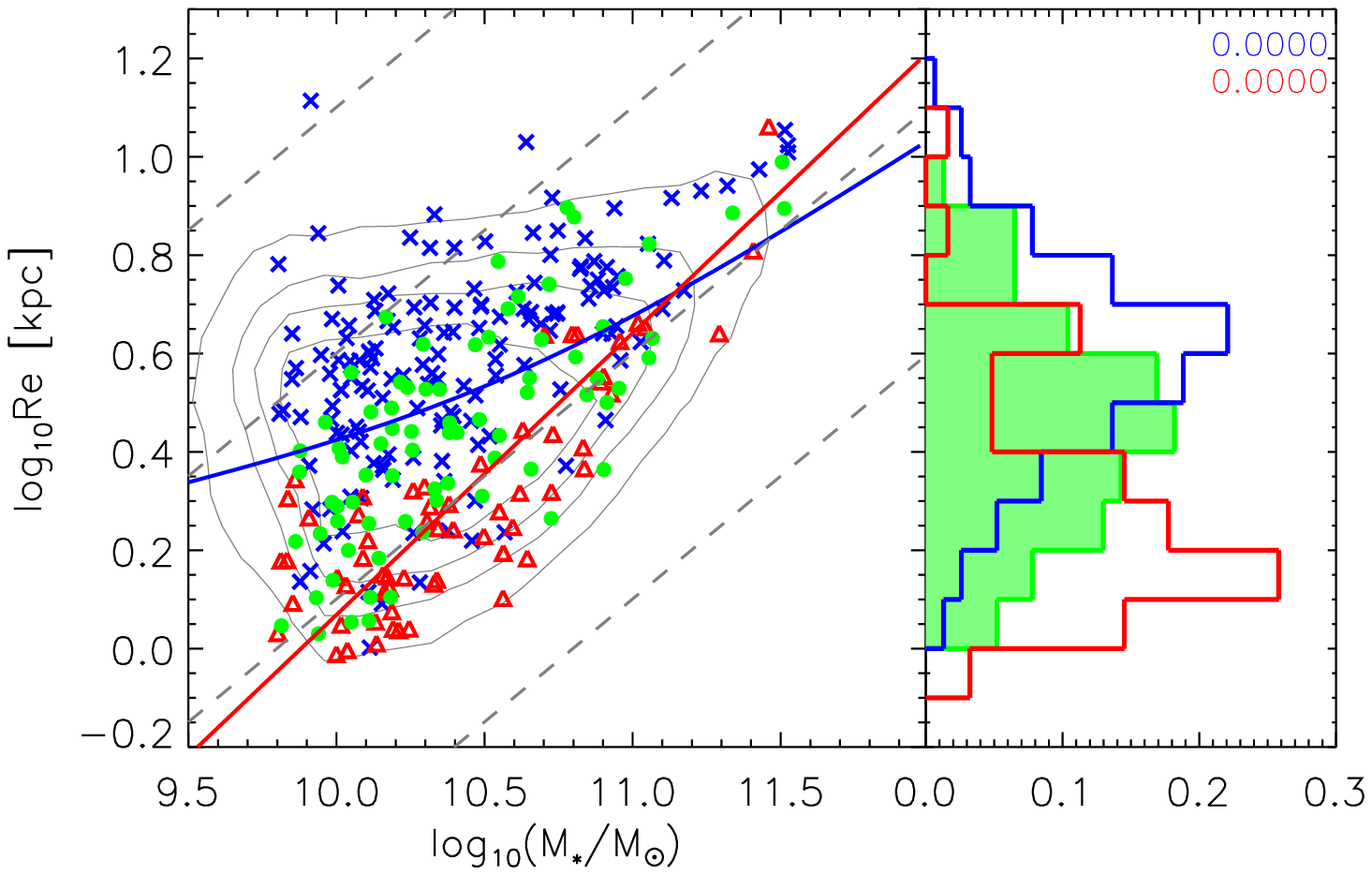,clip=true,width=0.48\textwidth}
    \epsfig{figure=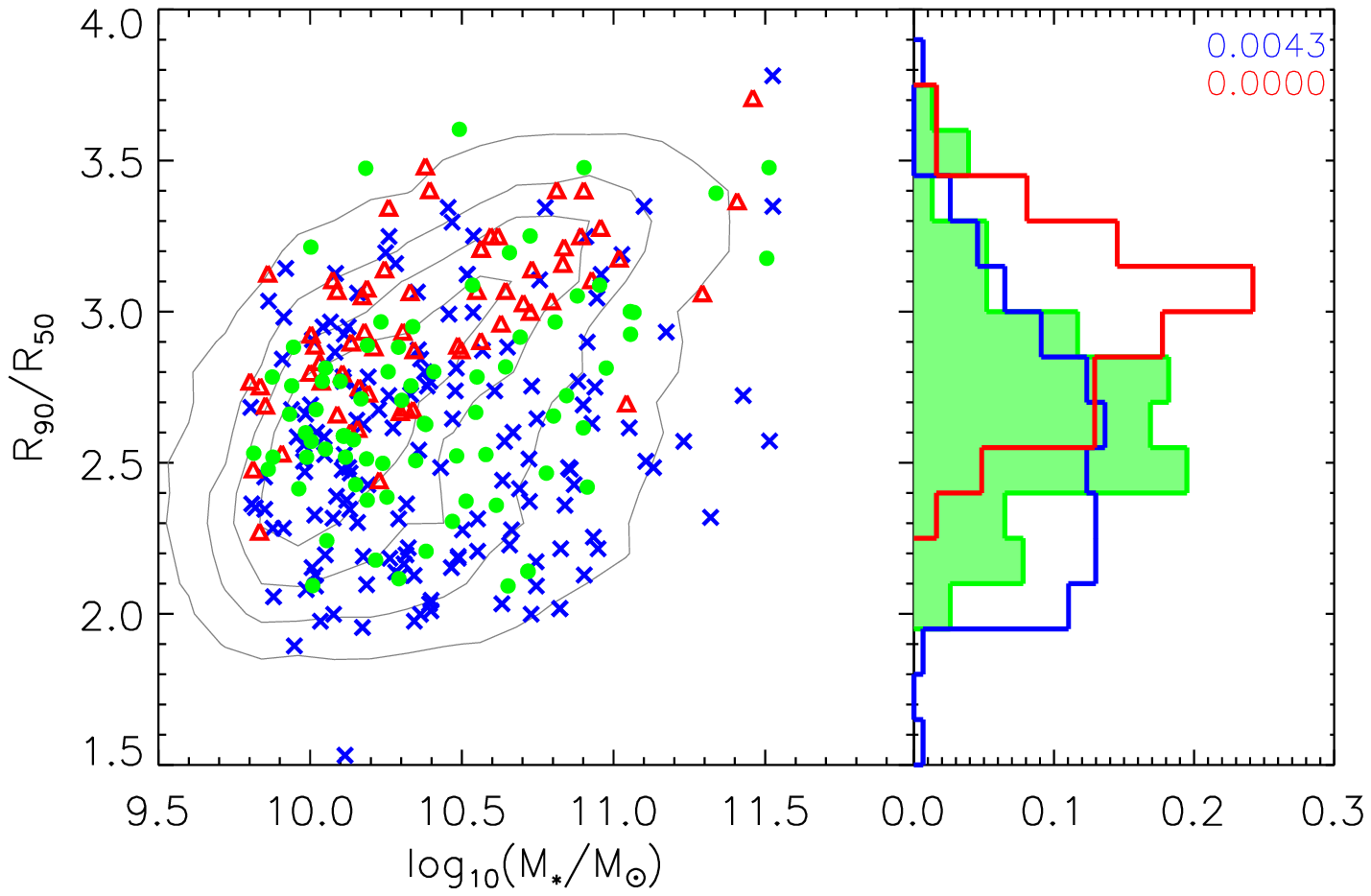,clip=true,width=0.48\textwidth}
    \epsfig{figure=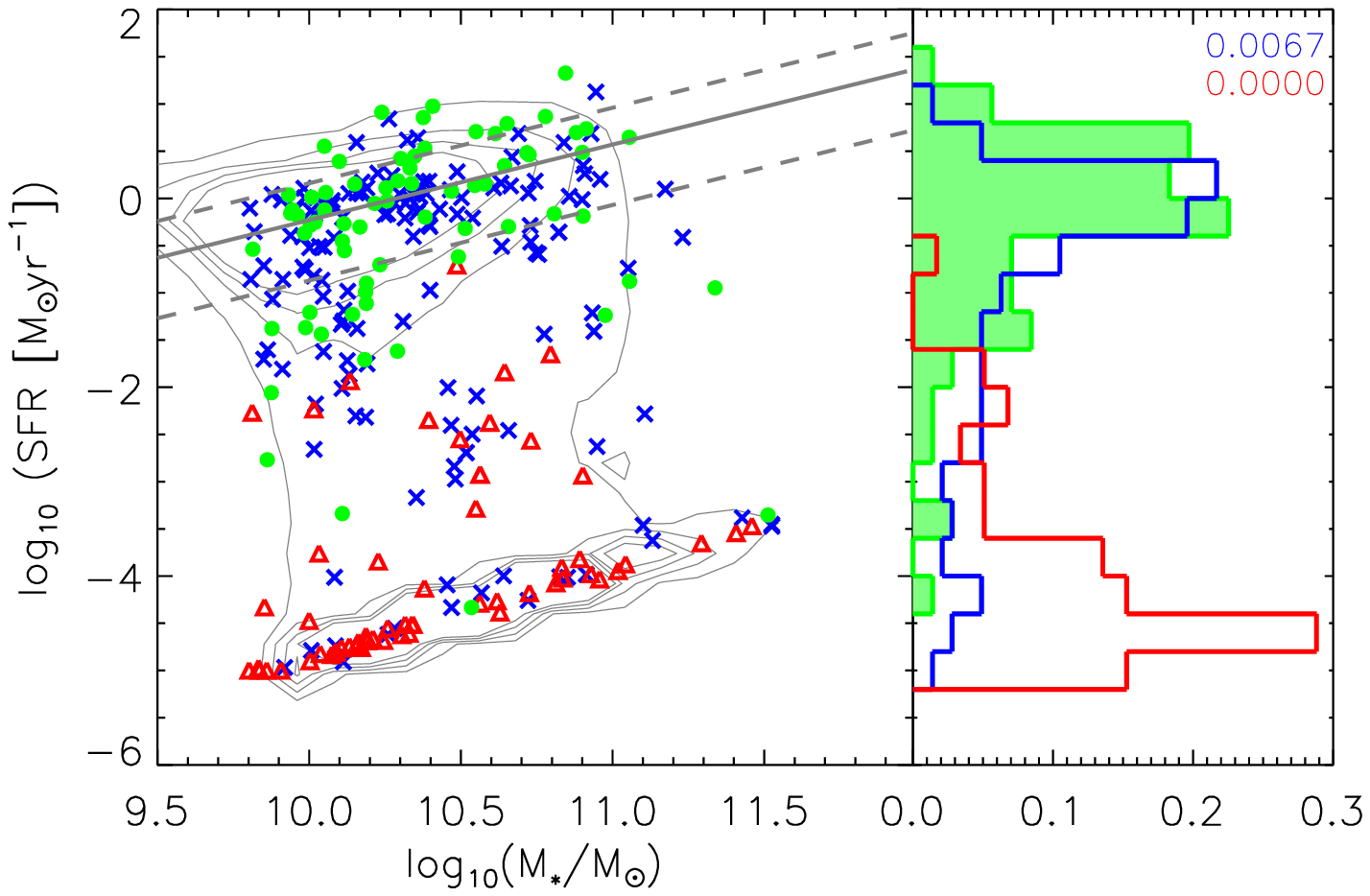,clip=true,width=0.48\textwidth}
    \epsfig{figure=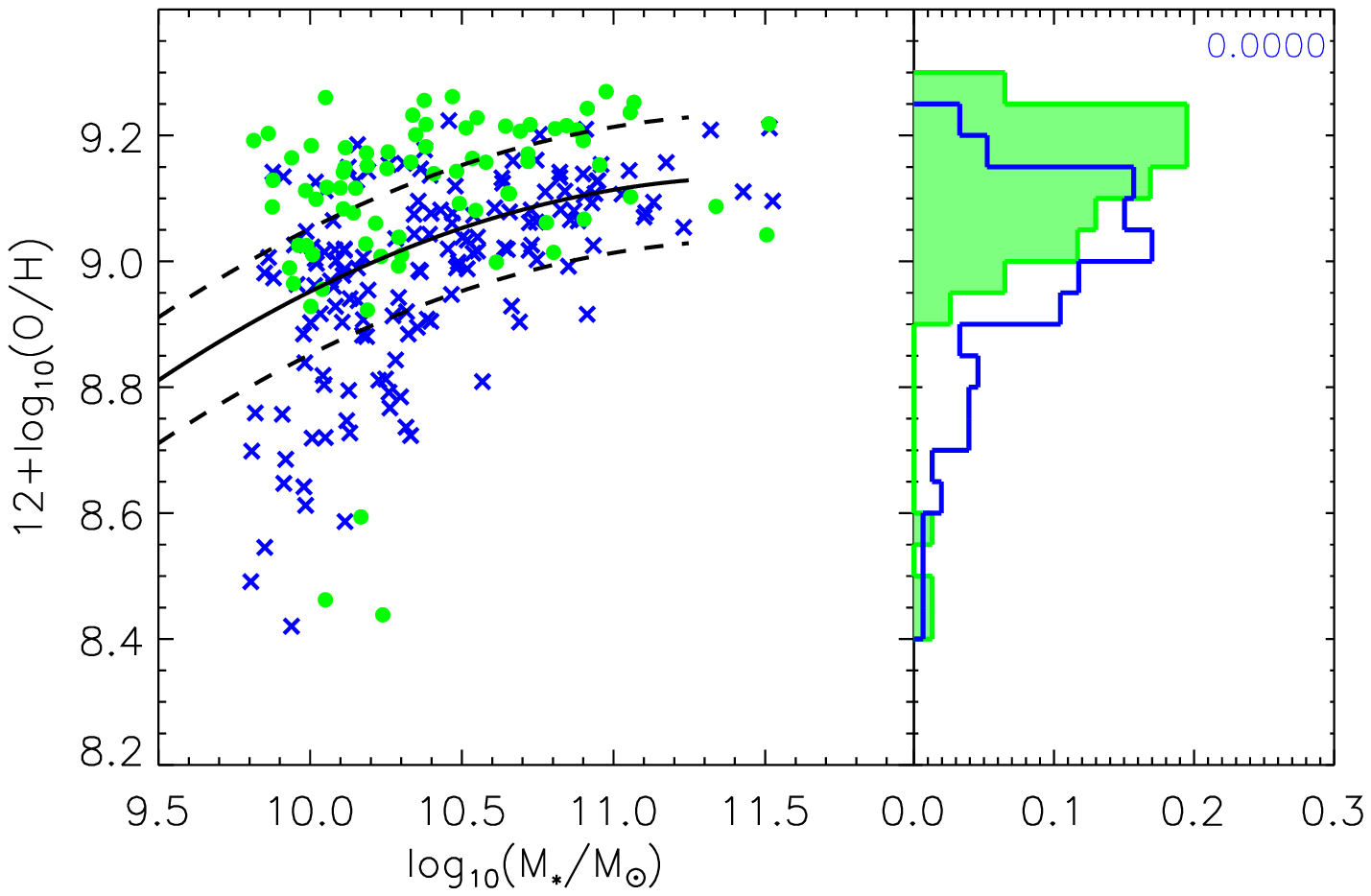,clip=true,width=0.48\textwidth}
  \end{center}
  \caption{The comparison of global properties for the targeted sample and two control samples.
  These four panels show the stellar mass versus size, concentration, star formation rate and 
  gas-phase metallicity for targeted sample (green dots), BC sample (blue crosses) and RC
   sample (red triangles), respectively. We also present the distribution of these four parameters for these three samples. 
   The K-S test probabilities for the difference in size, concentration, SFR and 
   gas-phase metallicity between targeted sample and BC/RC samples are denoted at the
   top-right corner. 
   For comparison, a volume-limited sample selected from SDSS is shown in the background contours in 
   all panels except the bottom-left one. 
   In the top-left panel, the blue and red solid curves represent the mass-size relation for late-type galaxies 
   (\sersic\ n $<$2.6) and early-type galaxies (\sersic\ n $>$2.6), taken from \cite{Shen-03}.
    The gray dashed lines represent constant surface mass density $\Sigma_e=$ 10$^7$, 10$^8$, 10$^9$ 
    and 10$^{10}$ M$_{\odot}$ kpc$^{-2}$ from top-left to bottom-right.
   In the bottom-left panel, the gray solid line shows the star formation main sequence taken from 
   \cite{Chang-15}, and the dashed lines show the 14\%-84\% region. 
   In the bottom-right panel, the black solid line shows the mass-metallicity relation taken from \cite{Tremonti-04}, and 
   the dashed lines show the 1$\sigma$ region.}
  \label{fig:global_properties}
\end{figure*}

In the previous section, we have selected the galaxies with recent outside-in  mass assembly mode.  
Then questions arise: whether their global properties, such as size, structural properties, global SFR and 
metal abundance, are different from control galaxies or not?
 
Figure \ref{fig:global_properties} shows the comparisons of global properties between targeted sample and two control samples. 
These four panels show the stellar mass versus size, concentration (\concen), 
star formation rate (SFR) and gas-phase metallicity for targeted sample (green dots), 
BC sample (blue crosses) and RC sample (red triangles), respectively.  
   In the bottom-right panel, only targeted sample and BC sample are shown,
   because the gas-phase metallicity are computed by using emission lines and red control galaxies usually 
   do not have prominent emission lines.  
    In the top-left panel, the blue and red solid curves represent the mass-size relation for late-type galaxies 
   (\sersic\ n $<$2.6) and early-type galaxies (\sersic\ n $>$2.6), taken from \cite{Shen-03}. 
    The gray dashed lines represent constant stellar mass surface density $\Sigma_e=$ 10$^7$, 10$^8$, 10$^9$ 
    and 10$^{10}$ M$_{\odot}$ kpc$^{-2}$ from top-left to bottom-right.
   In the bottom-left panel, the gray solid line shows the star formation main sequence taken from 
   \cite{Chang-15}, and the dashed lines show the 14\%-84\% region. 
   In the bottom-right panel, the black solid line shows the mass-metallicity relation taken from \cite{Tremonti-04}, 
   and the dashed lines show the 1$\sigma$ region.
   
  For comparison, a volume-limited sample selected from NASA-Sloan Atlas \citep[NSA;][]{Blanton-11} 
  catalog is shown in the background contours in all panels except the bottom-right one. 
  The volume-limited sample was selected to have $r$-band magnitude brighter than $-19.5$ and 
  redshift between 0.01 and 0.055. 
  We also present the distributions of these four parameters for these three samples (except for
   the gas-phase metallicity distribution of RC sample). 
 According to the distributions shown in Figure \ref{fig:global_properties}, we find significant differences
 between the size, concentration, SFR and gas-phase metallicity distributions of targeted 
 sample and two control samples. 
 We perform Kolmogorov-Smirnov (K-S) tests to quantify the significance, and the K-S test probabilities are 
 denoted at the top-right corner in each panel. The blue one is for the K-S probability
 between targeted sample and BC sample, and the red one is for the K-S probability
  between targeted sample and RC sample. 
             
  The stellar masses and sizes in Figure \ref{fig:global_properties} are from NSA catalog. 
  The size is the half light radius of SDSS $r$-band image. The concentration is defined as
   $R_{90}/R_{50}$, where $R_{90}$ and $R_{50}$
  are corresponding to the radii contain 90\%  and 50\% of light in SDSS $r$-band. 
    The SFR is the global star formation rate from \cite{Chang-15}, which is computed by the 
    SED fittings with SDSS and WISE broad-band fluxes. 
   
 The mass-size relation of galaxies has been well investigated in many recent literatures
 \citep[e.g.][]{Shen-03, McIntosh-05, Cappellari-13a}. SF galaxies 
 and quiescent galaxies have different mass-size relation, suggesting the galaxy structure 
 evolution along star formation cessation. 
 In the top two panels of Figure \ref{fig:global_properties}, the targeted galaxies broadly lie between the two
 control samples. 
 This indicates that galaxies with outside-in assembly mode have the structure properties 
 between that of normal SF galaxies and quiescent galaxies. 
 We have checked the fraction of smooth- and disk-like galaxies in 
 targeted sample and two control samples by using the data from Galaxy Zoo 2 \citep{Willett-13b}. 
 We find an increasing fraction of the smooth-like galaxies from the BC sample (36\%), 
 the targeted sample (54\%) to the RC sample (86\%). In addition, 
 the targeted galaxies have a median bulge fraction \citep{Simard-11} of 0.36, 
 which is higher than that of BC galaxies (0.26), and lower than that of RC galaxies (0.57). 
 These agree well with the results in top two panels of Figure \ref{fig:global_properties}. 
 
The bottom-left panel shows that the targeted galaxies tend to have higher global SFR than BC galaxies
for about 0.30 dex. The majority of targeted galaxies with \nuvr$<$4 are above the median star 
formation main sequence from \cite{Chang-15}.  
This is contradict with the general knowledge of galaxies that more concentrated galaxies tend to be less 
active in forming stars \citep{Bell-12, Fang-13, Bluck-14, Barro-17}, suggesting that the targeted 
galaxies are indeed an unusual population with respect to normal SF galaxies. 
 
The mass-metallicity relation is one of the basic relations in galaxy formation and evolution field, 
which has been studied in many works \citep[e.g.][]{Lequeux-79, Garnett-Shields-87, Tremonti-04, Gallazzi-05, 
Panter-08, Sanchez-13, Lian-15}. 
The bottom-right panel shows that the targeted galaxies are more metal-rich than BC galaxies. 
The oxygen abundance used in this panel is the H$\alpha$ flux weighted metallicity for all spaxels based on 
MaNGA data. For each spaxel, the oxygen abundance is calculated using N2O2 indicator, with adopting 
$\rm log_{10}(O/H)+12= log_{10}[1.54020 +1.26602\times R + 0.167977\times R^2]+8.93$, 
where $\rm R=log_{10}[NII]/[OII]$ \citep{Dopita-13}. \cite{Zhang-17} proposed that N2O2 indicator is optimal in 
measuring metallicity for diffused ionized regions, because it is not sensitive to ionization parameter \citep{Dopita-00,Dopita-13, Kewley-Dopita-02}. 
As shown in this panel, the median gas-phase metallicity of targeted sample is about 0.12 dex
higher than that of BC galaxies. 

There are three targeted galaxies with dramatically low gas-phase metallicities ($\rm 12+log_{10}(O/H)<8.6$), 
which is about 0.4 dex lower than the median relation from \cite{Tremonti-04}. One of them is 
an edge-on galaxy and the other two show blue clumpy structures in SDSS color image. 
Their low gas-phase metallicities suggest an external gas acquisition. 
 In contrast, the cold gas in other targeted galaxies, as fuels of star
 formation, may be pre-existing gas, and the metal abundance is later populated by the recent 
 star formation \citep{Chen-16}. 

\subsection{2D maps and radial profiles of three diagnostic parameters}
\label{subsec:SFH_profiles}

\begin{figure*}
\rotatebox{270}{\epsfig{figure=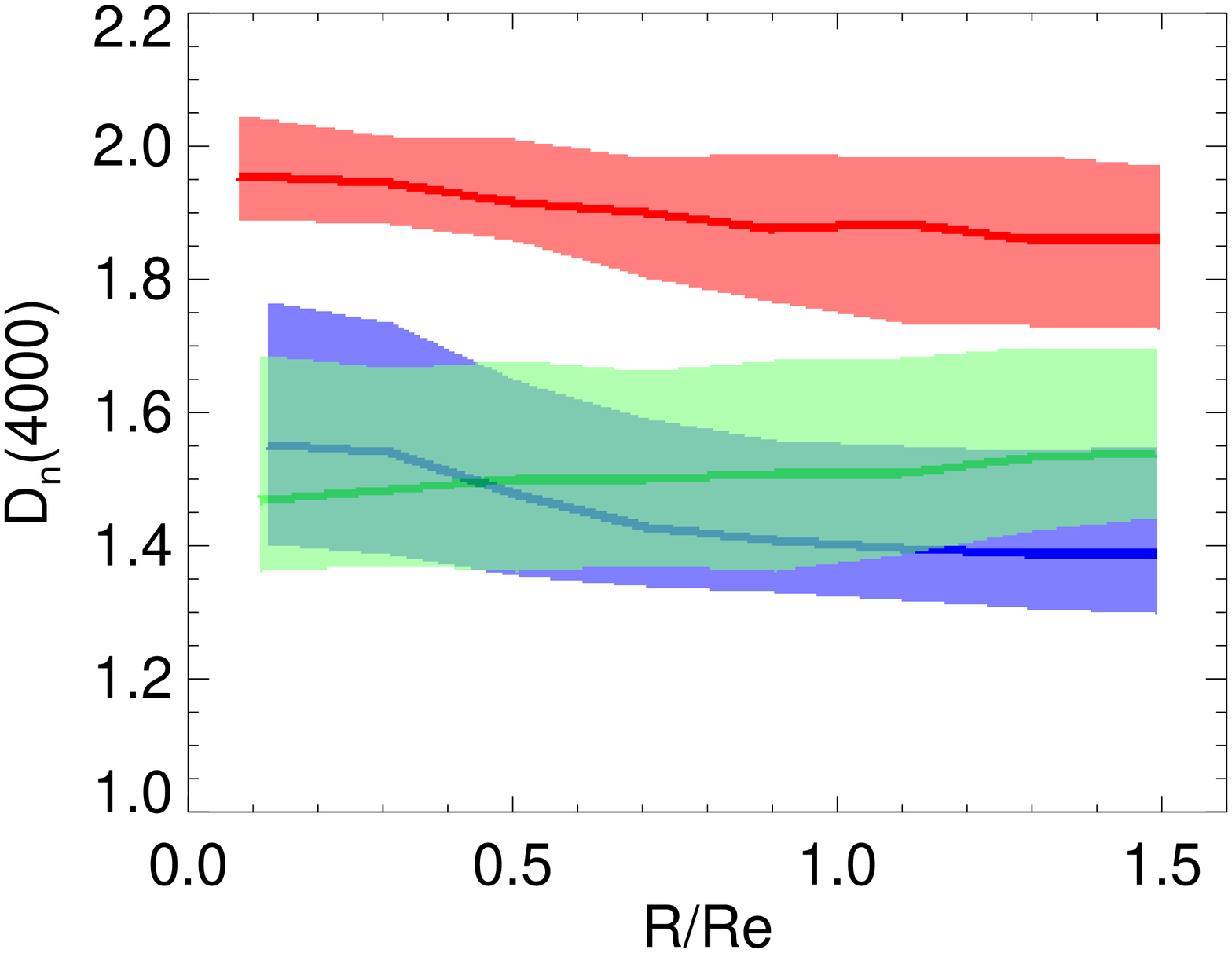,clip=true,width=0.25\textwidth}}
\rotatebox{270}{\epsfig{figure=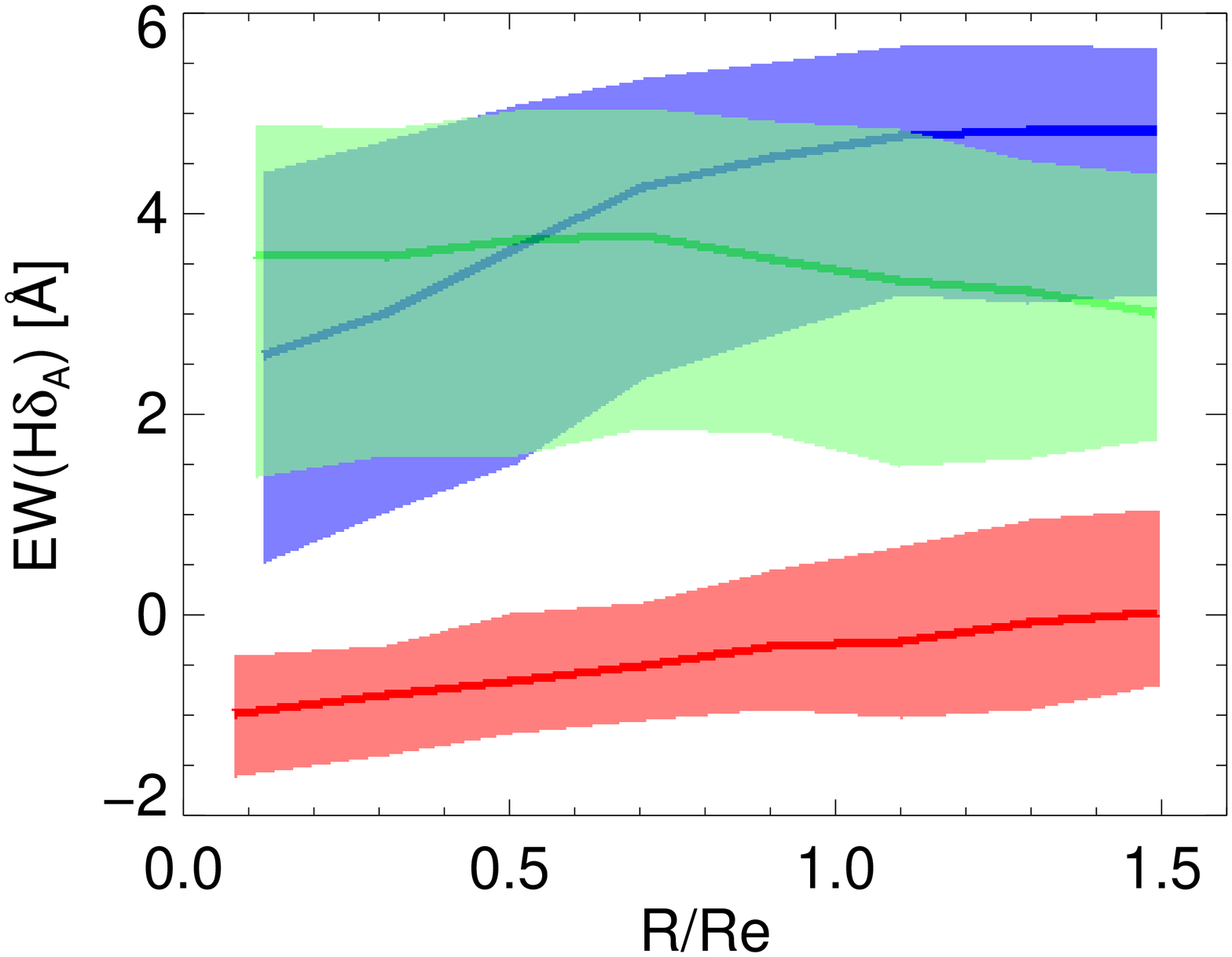,clip=true,width=0.25\textwidth}}
 \rotatebox{270}{\epsfig{figure=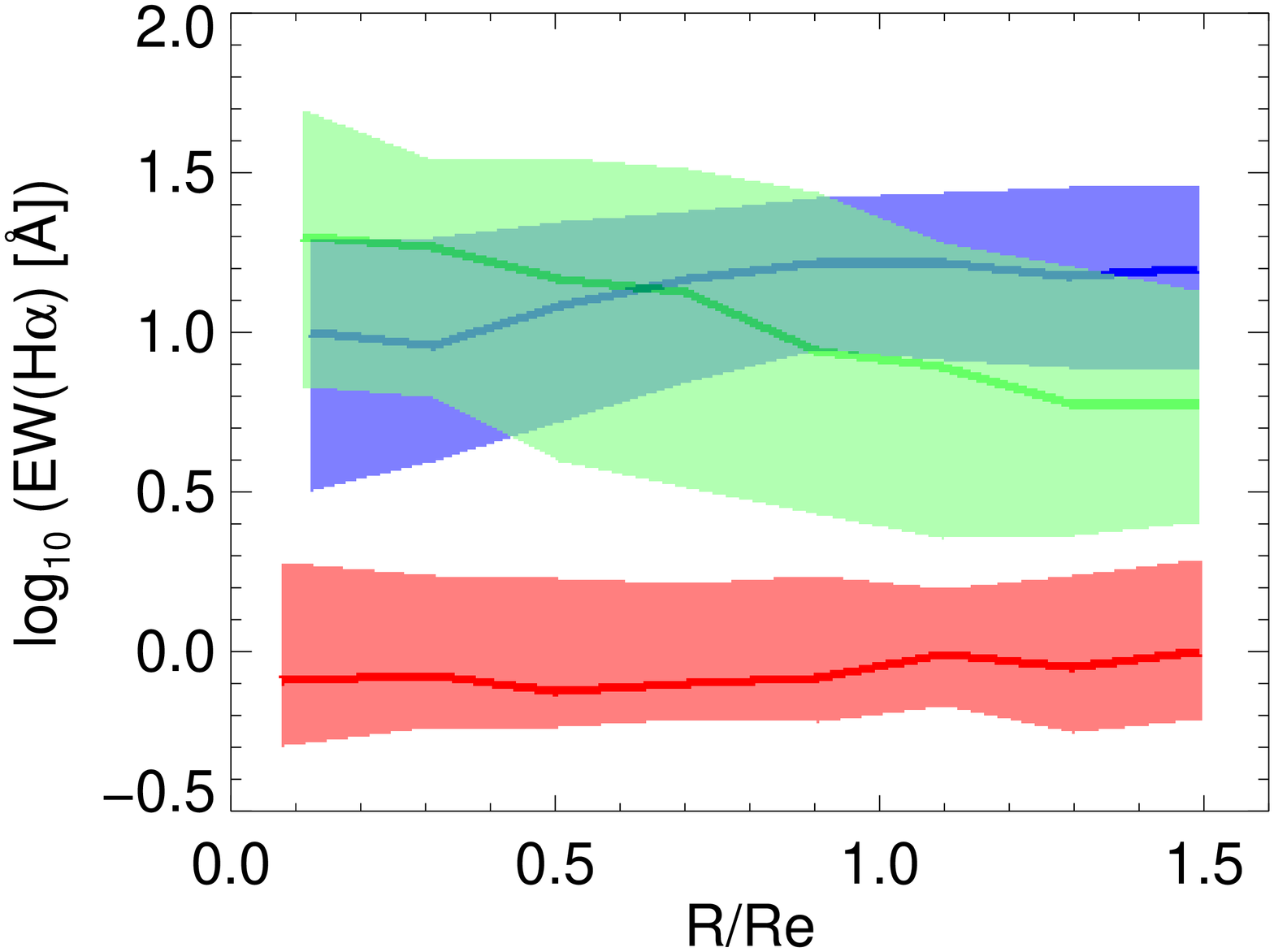,clip=true,width=0.25\textwidth}}
\caption{Radial profiles of 4000 \AA\ break (left panel), \hda\ absorption index (middle panel) 
and \lgewhae\ (right panel) for our targeted sample and two control samples. 
In each panel, the blue, green and red curves show the median profile for
 BC sample, targeted sample and RC sample, respectively.  
 The shaded regions are corresponding to the 20-80 per cent percentile range. 
 We present the radial profile with radii scaled the effective radius. }
 \label{fig:SFH_profile}
\end{figure*}

In addition to the global properties investigated in the previous section, the IFS survey enables us to study 
the resolved properties of our samples.
Figure \ref{fig:SFH_profile} shows the radial profiles of the 4000 \AA\ break (left panel), \hda\ index (middle panel) 
and \lgewhae\ (right panel) for these three samples.
In each panel, the solid blue, green and red curves show the median profile for BC sample, targeted sample and
RC sample, respectively. The shaded regions are corresponding to the 20-80 per cent percentile range.
In presenting the radial profile, we scale the radii with the effective radius of SDSS $r$-band. 

For an individual galaxy, the radial profile of the 4000 \AA\ break is computed in the following way.  
We separate all spaxels with the continuum SNR$>$5 at 5500 \AA\ into a set of radial intervals with 
a constant width of $\rm \Delta log_{10}(R/Re) = 0.2$. 
At a given radial bin for an individual galaxy, the value of 4000 \AA\ break profile is estimated by median 
\dindex\ of the spaxels falling in that radial bin. 
In this process, we have corrected the inclination effect for each spaxel based on the minor-to-major axis ratio 
from NSA \citep{Blanton-11}. Throughout the paper, 
we calculate the radial profiles in the way above for all referred parameters.

As expected, the targeted and BC samples have lower \dindex, higher \ewhda\ and higher \lgewhae\ than RC
 sample as a whole. The BC sample has negative \dindex\ gradients, and positive \ewhda\ 
 and  \lgewhae\ gradients within the effective radius. The radial profiles of BC sample 
become flat when the radii greater than \re.  
In contrast, the radial profiles of targeted sample show 
slightly positive gradients of \dindex, slightly negative gradients of \ewhda\ and 
pronounced negative gradients of \lgewhae. The gradients 
of these three parameters still exist when the radius is beyond \re.
Figure \ref{fig:SFH_profile} confirms that the targeted galaxies have stronger star formation 
activities in the centers than the outer regions. 
This is opposite to the previously findings of inside-out assembly mode in massive 
SF galaxies \citep[e.g.][]{Munoz-Mateos-07, Bezanson-09, Perez-13, Pan-15, Tacchella-15}.
The targeted galaxies are very interesting, since they are associated with rapid central stellar mass
assembly (or bulge growth).  Investigating their properties would probably give clues for the processes 
of bulge growth and star formation cessation (More will be discussed in Section \ref{subsec:quenching}). 

\subsection{Radial profiles of metallicity and stellar surface mass density}
\label{subsec:metal_profiles}

\begin{figure*}
  \begin{center}
    \rotatebox{270}{\epsfig{figure=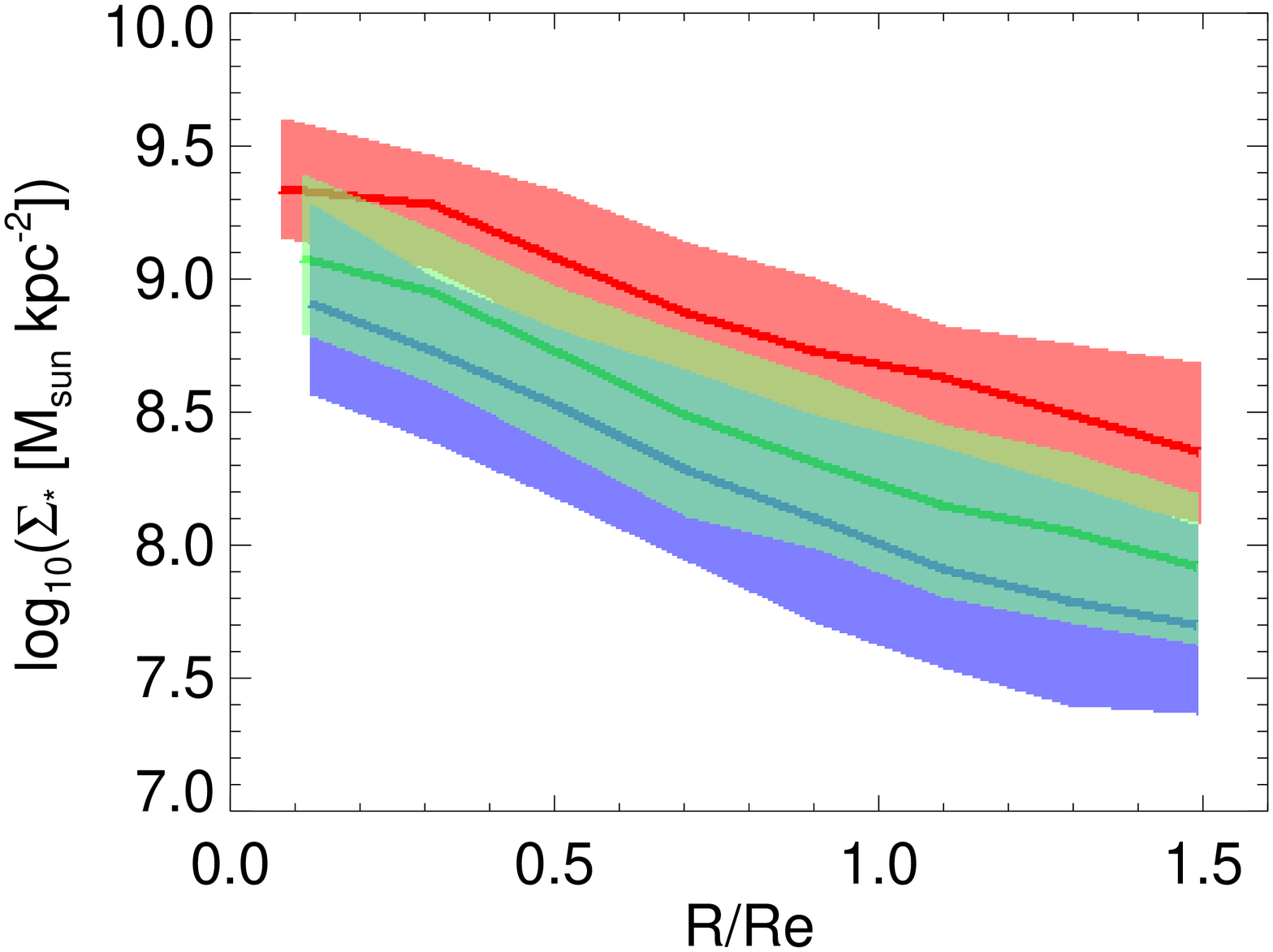,clip=true,width=0.30\textwidth}}
    \rotatebox{270}{\epsfig{figure=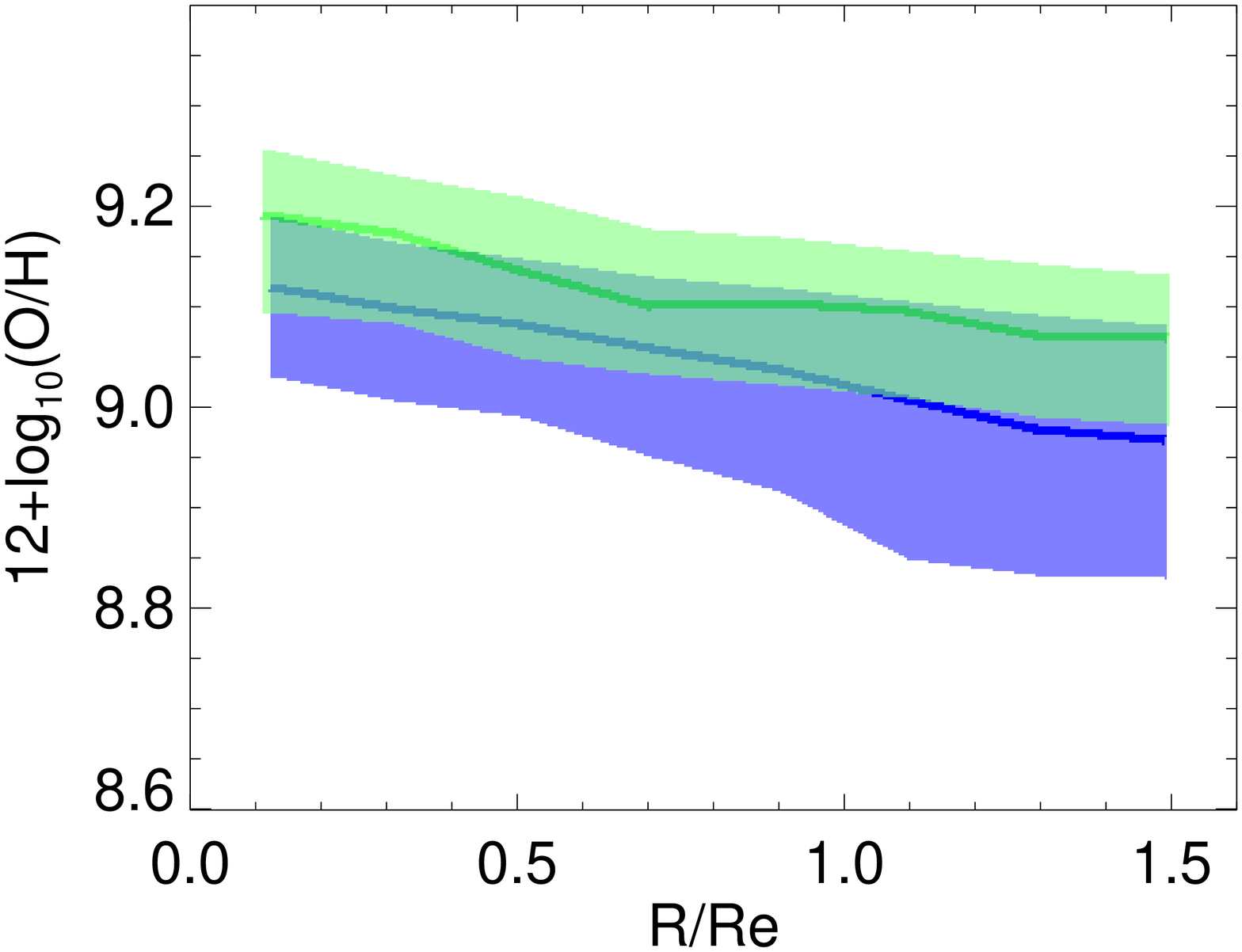,clip=true,width=0.30\textwidth}}
  \end{center}
  \caption{Radial profiles of stellar mass surface density (left panel) and gas-phase metallicity (right panel).
  In the left panel, the blue, green and red curves represent the stellar surface density profiles of  
  BC sample, targeted sample and RC sample. In the right panel, we only present the
   gas-phase metallicity for BC sample (blue) and targeted sample (green). 
 The shaded regions are corresponding to the 20-80 per cent percentile range. 
   The radial profiles are shown with the radii scaled the effective radius.}
 \label{fig:MZ1}
\end{figure*}

\begin{figure}
  \begin{center}
    \epsfig{figure=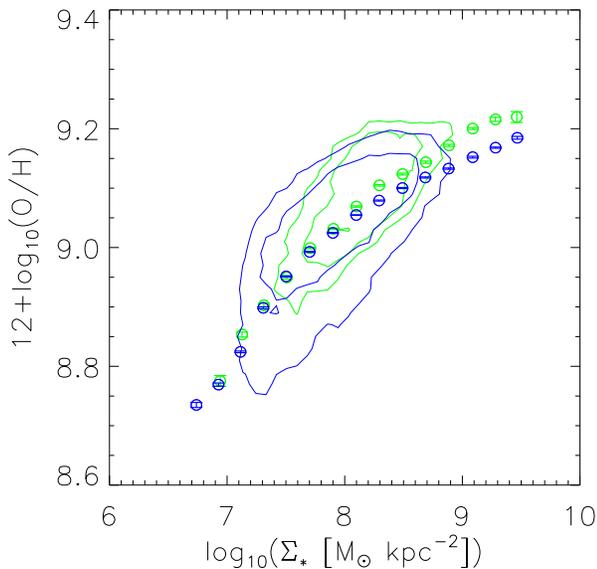,clip=true,width=0.45\textwidth}
  \end{center}
  \caption{The $\Sigma_*$-$Z_{gas}$ relation for star formation regions in targeted galaxies (green contours) and
   BC galaxies (blue contours). The blue and green data points represent the median relations. 
   The errors are computed using bootstrap method. }
 \label{fig:MZ2}
\end{figure}

Figure \ref{fig:MZ1} shows the median radial profiles of stellar surface mass density ($\Sigma_*$, left panel) 
and gas-phase metallicity (right panel) for our samples. The shaded regions are corresponding to the 20-80 per cent 
percentile range. The stellar mass maps are the outputs from the STARLIGHT fittings (see Section \ref{subsec:spectral_fitting}). 
As shown in the left panel of Figure \ref{fig:MZ1}, the RC sample has the highest stellar surface density and the
BC sample has the lowest stellar surface density as a whole. 
The median $\Sigma_*$ profiles of three samples are nearly parallel in logarithmic space. 
The targeted sample has higher median $\Sigma_*$ profile than BC sample for about 0.22 dex, while 
lower than RC sample for about 0.39 dex. 
This result is consistent with the results in Figure \ref{fig:global_properties}, indicating that 
the structural properties of targeted sample are between the two control samples. 

The right panel of Figure \ref{fig:MZ1} shows the gas-phase metallicity profile for BC 
sample (blue line) and targeted sample (green line). 
As shown, the targeted sample is systematically more metal-rich than BC sample 
for $\sim$0.07 dex within 1.5\re. 
The median metallicity profile of BC sample is slightly steeper than targeted sample. 
The BC sample has a median metallicity gradient of -0.11 dex/\re, 
which is slightly steeper than that of targeted sample (-0.08 dex/\re). 
The gradients measured in this work are comparable with the previously measurements of -0.1 dex/\re\  
by \cite{Sanchez-14}.

Recently, the $\Sigma_*$ and local gas-phase metallicity are found to be closely related, with the
 higher $\Sigma_*$ the higher metallicity \citep{Barrera-Ballesteros-16, Zhu-17}.
Thus one may ask the question: whether the higher metallicity of targeted sample is only due to the higher 
stellar surface density or not?  To answer this question, we plot the $\rm log_{10}\Sigma_*$ vs. metallicity for
 all spaxels in BC sample (blue contours) and targeted sample (green contours) in Figure \ref{fig:MZ2}, 
 excluding the spaxels with the SNR of [OII]$\lambda$3727 flux, H$\beta$ flux, 
H$\alpha$ flux, or [NII]$\lambda$6584 flux less than 3.  
In Figure \ref{fig:MZ2}, the blue and green data points represent the median relations for BC sample and
 targeted sample, respectively. The targeted galaxies are more concentrated than BC galaxies, 
and have larger stellar surface density as a whole.
The targeted sample is more metal-rich than the BC sample 
at high $\Sigma_*$ end ($\Sigma_*>$10$^{8.5}$ M$_{\odot}$ kpc$^{-2}$).
The metallicities of these two samples become comparable when $\Sigma_*<$10$^8$ M$_{\odot}$ kpc$^{-2}$. 
 We find that the inner region of targeted galaxies are more metal-rich than the inner region of
 BC galaxies even at fixed stellar surface mass density. 
 With respect to BC galaxies, the higher central oxygen abundance in targeted galaxies
 may be due to the enhancement from the on-going intense star formation activities as proposed
  by \cite{Chen-16}. By using highly resolved IFS data of HGC 91c, 
  \cite{Vogt-17} found the metal enhancement has proceeded preferentially along spiral arms, and less 
  efficiently across the arms, suggesting the metal enhancement and intense star formation activities association. 

\subsection{Environment of our samples}
\label{subsec:environments}

We have found that the structural properties and radial star formation properties of targeted sample are significantly
distinct from the BC sample. In this section, we investigate the environment of our samples, and try to 
find out whether they reside in the same environment or not. 

We use the projected two-point cross-correlation function (2PCCF, $w_p(r_p)$) to quantify the clustering 
properties of our samples. Here we briefly describe the method to calculate 2PCCFs,
 and a full description of the methodology for computing 2PCCF was presented in \cite{Li-06}. 
The reference sample is from the version dr72 of the NYU-VAGC: 
a spectroscopic reference sample. It is a magnitude-limited sample of about half a million galaxies with $r$-band
Petrosian apparent magnitude corrected for galactic extinction $r<$17.6, the absolute magnitude range of
$-23<M_{0.1r}<-17$, and the redshifts in the range of 0.01$<z<$0.2. Here, M$_{0.1r}$ is the $r$-band 
Petrosian absolute magnitude, corrected for evolution and K-corrected to its value at $z=0.1$. 
In calculating 2PCCF, we have generated a random sample that has the same selection effects as the 
reference sample \citep{Li-06, Lin-14}. 
 We cross-correlate the targeted sample as well as two control samples, with respect to the reference sample and 
 the random sample, then define $w_p(r_p)$ as the ratio of the two pair counts minus one in the projected 
 separation $r_p$. The effect of fiber collisions is carefully corrected following the method in \cite{Li-06}. 
 
\begin{figure}
  \begin{center}
    \epsfig{figure=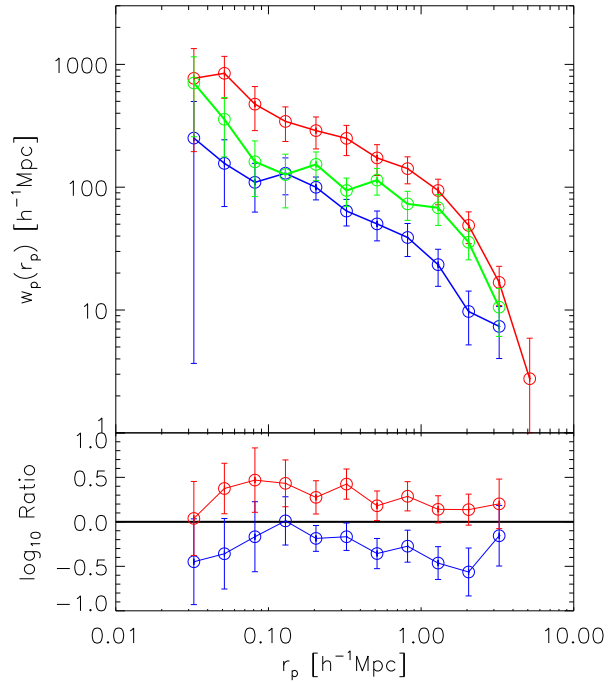,clip=true,width=0.45\textwidth}
  \end{center}
  \caption{Projected two point correlation function for targeted sample ($w_p^{\rm T}(r_p)$, green curve), 
   BC sample ($w_p^{\rm BC}(r_p)$, blue curve) and RC sample ($w_p^{\rm RC}(r_p)$, red curve).
    The amplitude ratios, $w_p^{\rm BC}(r_p)$/$w_p^{\rm T}(r_p)$ and $w_p^{\rm RC}(r_p)$/$w_p^{\rm T}(r_p)$, are presented 
    in the bottom panel. The errors are computed by using the bootstrap resampling technique.
 }
 \label{fig:2PCF}
\end{figure}

The top panel of Figure \ref{fig:2PCF} shows the 2PCCFs measured for BC sample ($w_p^{\rm BC}$, blue curve), 
targeted sample ($w_p^{\rm T}$, green curve) and RC sample ($w^{\rm RC}_p$, red curve).
For clarity, we also include the amplitude ratio, $w_p^{\rm BC}/w_p^{\rm T}$ and $w_p^{\rm RC}/w_p^{\rm T}$, in logarithmic space 
in the bottom panel. 
The errors in the $w_p(r_p)$ measurements are computed by using the bootstrap
resampling technique \citep{Barrow-Bhavsar-Sonoda-84}. The data points of $w_p(r_p)$ are not shown 
on scales smaller than 30 kpc, because 
they have very large uncertainties due to the limitation of sample sizes. 

As expected, the RC galaxies are more strongly clustered than BC galaxies at a large 
range of scales from 30 kpc to 3 Mpc. As a whole, the clustering amplitude of targeted sample 
is between that of BC sample and RC sample. 
The targeted galaxies are more strongly clustered than BC galaxies almost at the whole scale we considered, 
especially at the scale of 0.8-2 Mpc (0.3-0.5 dex higher). 
However, the targeted sample appears to be more weakly clustered than 
RC sample, especially at the scale of  50-300 kpc (0.3-0.5 dex lower). 
Assuming the environment effect is the main driver for the outside-in mode in targeted galaxies,  
it is difficult to understand why the environment at large scale ($\sim$ 1 Mpc)  could have 
an effect on the SF properties of galaxies.
At the scale of 30 kpc, the targeted sample appears to be more strongly clustered than BC sample.  
However, we do not overinterpret the result 
in Figure \ref{fig:2PCF} because of the limitation of sample size.



In addition to 2PCCFs, we also use the three-dimensional reconstructed local mass density ($\delta=\rho/\bar{\rho}$, where 
$\rho$ is the local matter density and $\bar{\rho}$ is the mean matter density) 
to quantify the environment for each galaxy in our sample. 
The mass density field is taken from the Exploring the Local Universe with ReConstructed Initial Density Field
project \citep[ELUCID;][]{Wang-16}, and matches well the galaxy and group distribution \citep{Yang-07}
of the SDSS Data Release 7 \citep{Abazajian-09}. 
The local mass density we used is the smoothed density by a Gaussian kernel with a scale of 2 Mpc h$^{-1}$ 
at each galaxy position.  
     
\begin{figure}
  \begin{center}
   \epsfig{figure=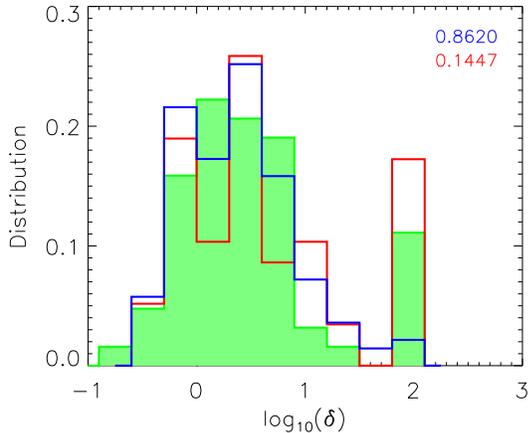,clip=true,width=0.45\textwidth}
  \end{center}
  \caption{Density distributions of BC sample (blue histogram), targeted sample (green filled
   histogram) and RC sample (red histogram). The K-S test probability of targeted sample and 
   BC sample is denoted in blue, and the K-S test probability of targeted sample and RC sample is denoted in red. }
  \label{fig:overdensity}
\end{figure}

Figure \ref{fig:overdensity} shows the density distribution of BC sample (blue histogram), targeted sample (green 
filled histogram) and RC sample (red histogram). 
All of these three samples have broad density distributions. 
To quantify the significance of the distribution difference between targeted sample and two control 
samples, we present the K-S test probabilities at the top-right corner. 
The K-S test probabilities suggest that significant differences between the density distribution
of targeted sample and two control samples are not found.
We note that about 17\% of RC galaxies, 11\% of targeted galaxies and 2\% of BC galaxies 
reside in the dense regions with $\rho/\bar{\rho}\sim$100.    

The results of Figure \ref{fig:2PCF} and Figure \ref{fig:overdensity} bring to a question: 
whether the large distinction of clustering amplitude between targeted and BC samples at
scale of 0.8-2 Mpc is mainly contributed by the 11\% of targeted galaxies resided in big clusters?
To answer this question, we have checked the clustering properties of targeted and BC samples 
with excluding the galaxies resided in the most massive halos (M$_h>10^{14}$\msun). 
The result shows that the clustering amplitude of targeted 
galaxies has been reduced to be slightly higher than (or comparable with if considering the errors) 
BC sample at scale of 0.8-2 Mpc. 
However, the targeted galaxies still appear to be more strongly clustered than BC sample
 at the scale of 30 kpc. 
 
Further more, we have calculated the projected separation of the nearest neighbour for the targeted and BC galaxies 
with the velocity differences less than 500 km s$^{-1}$.  In this process, we ignore the spectroscopic incompleteness 
effect, since we just care about the comparison of companion fraction between targeted and BC galaxies. 
We find that 19 ($\sim$25\%) targeted galaxies and 
14 ($\sim$10\%) BC galaxies have one or more companions within 100 kpc. 
This suggests interaction with close companions is likely to be one of the drivers for the outside-in assembly mode.  Since the majority of targeted galaxies do not have companions within 100 kpc,
 it appears that this mechanism is not the main driver. 
We will discuss the possible mechanisms for the formation of the targeted galaxies in Section \ref{subsec:formation}.
   

\section{Discussion}
\label{sec:discussion}


\subsection{The properties of targeted sample}
\label{subsec:discussion1}

The IFS survey enables us to select a sample of massive SF galaxies with 
recent outside-in stellar mass assembly mode. 
They are selected to have \dindexout$>$\dindexin, and account 
for 14\% of the total massive SF galaxies (\nuvr$<$5 and \mstar$>10^{9.8}$\msun).  
For comparison, we have built two control samples matched in stellar mass. 
On average, the targeted galaxies appear to have smaller sizes, higher concentration, 
higher SFR and higher oxygen abundance than normal SF galaxies.

We have generated the \dindex, \ewhda\ and \lgewhae\ maps and profiles for our samples. 
The inside-out mass assembly scenario has been found in observations for massive SF
 galaxies \citep[e.g.][]{Sanchez-Blazquez-07, Perez-13, Ibarra-Medel-16}.
However, the targeted galaxies appear to have significantly different 2-dimensional 
mass assembly pattern with normal SF galaxies. They have lower \dindex\ and  higher \lgewhae\ in their 
inner regions than in outer regions, suggesting that they host younger stellar populations 
and more intense star formation activities in inner regions than in outer regions. 
The higher metallicity in the inner regions of targeted galaxies are likely due to the enhancement of 
recent intense star formation activities.

\cite{Lin-17} have identified a sample of 17 ``turnover'' galaxies from 57 nearly face-on spiral CALIFA 
galaxies, whose central regions exhibt significant drop in \dindex.  The central regions of these galaxies have 
experienced star formation in the past 1-2 Gyr.  Nearly all the turnover galaxies are barred, 
suggesting a bar-induced star formation scenario in galaxy center \citep[e.g.][]{Hawarden-86, 
Ho-Filippenko-Sargent-97, Wang-12, Cole-14}. The targeted galaxies in this work 
is different from their population, because the \dindex\ profiles of targeted galaxies do not show turnover 
feature at galactic center. We have checked the bar fraction of the targeted galaxies, and find that about 
19\% of targeted galaxies appear to have bar-like structure \citep{Willett-13b}. 
This indicates that existence of the bar is not mainly responsible for the outside-in mass assembly
 mode in targeted galaxies.  

From 489 star forming galaxies of MaNGA sample, \cite{Chen-16} have identified 9 counter-rotators, whose 
gas and stars are kinematically misaligned. They also have positive slope of \dindex\ profiles, and more intense,
 on-going star formation in their central regions than in outer regions \citep{Chen-16, Jin-16}. 
Five of them are also selected in our targeted sample, and the other four galaxies have the stellar mass
 lower than our selection criteria. %
They proposed that the formation of SF counter-rotators is due to gas 
accretion from gas-rich dwarfs or cosmic web. 
The counter-rotators tend to reside in more isolated environment \citep{Jin-16}, while our targeted 
galaxies live in a variety of environments  (see Figure \ref{fig:overdensity}). 
This suggests that the gas accretion may be one of the mechanisms for the outside-in assembly mode, 
while it is not known whether it is the main driver.

\subsection{The formation of targeted galaxies}
\label{subsec:formation}

The targeted galaxies can be formed in two ways according to their abnormal resolved star formation 
properties.  One is that normal SF galaxies surfer from some kind of processes, such as interaction or 
gravitational instability of the disk, leading to gas inflow and triggering central star formation,
 and subsequently the galaxies with outside-in assembly mode are formed. 
The other is that quenched galaxies accrete cold gas
and re-form stars in their centers,  which can also lead to the observed outside-in assembly picture.
Figure \ref{fig:MZ1} shows that the central $\Sigma_*$ of targeted galaxies are between that of BC galaxies and 
RC galaxies. If the targeted galaxies are mainly formed via gas accretion of quenched galaxies, 
the central $\Sigma_*$ of targeted galaxies should be on average comparable with (or higher than) 
that of RC galaxies. Thus we deduce that the progenitors of the targeted galaxies are probably normal 
SF galaxies, which might later undergo cold gas inflow and morphological transformation. 

We find that 14 out of 73 targeted galaxies have bar-like structure 
(Four galaxies with inclination angle greater than 75 degree are not included in counting the bar fraction), 
and 19 out of 77 targeted galaxies have one or more companions within 100 kpc. 
It is well established that both the existence of the bar and interaction with
 companions can effectively cause the gas inflow and enhance star formation in galaxy center \citep{Wang-12,
 Li-08a}, although they do not always work \citep{Lin-17}.  
Among galaxies with companions, some of them are located in very massive clusters, which 
may suffer from a variety of environmental processes, such as tidal stripping, ram pressure stripping 
and galaxy harassment \citep[e.g.][]{Moore-96, Cox-06, Cheung-12}. 
 However in our case, both the environmental effect and existence of the bar
  are not enough to explain most of individual galaxies, suggesting that other mechanisms are needed.
 Specifically, there are 44 ($\sim$60\%) out of 73 targeted galaxies, who do not have 
 bar-like structure nor companions within 100 kpc.  
 Four of them are the counter-rotators in \cite{Chen-16}, which are proposed to undergo gas accretion from
  gas-rich dwarfs or cosmic web. 
 It is possible that more targeted galaxies are undergoing the same process, since the gas accretion do not always
 cause the kinematical misalignment of gas and stars.  
 
 Recently, in addition to mergers, an additional channel for bulge growth via disk instabilities was 
 proposed in semi-analytical model \citep{Porter-14b}, which brought the model into better agreement
  with the observed galaxy stellar mass function of spheroid-dominated galaxies. 
 Further, \cite{Brennan-15} confirmed that a model adding the disk instabilities for bulge growth 
 agrees better with the observed galaxy distribution on the sSFR-\sersic\ index plane than a model
 in which bulges grow only through mergers.  
 By implementing the bulge growth model with the mergers and disk instabilities into  the semi-analytic 
 galaxy formation code, \cite{Tonini-16} produced two distinct populations of bulges: merger-driven bulges, 
 similar to classical bulges and instability-driven bulges, similar to pseudo-bulges.  
The instability-driven bulges dominate the population of galaxies with intermediate stellar masses ($10<$\lgmstar$<11$), which are the similar mass range with our targeted galaxies.
Considering the very low major merger rate in local universe, 
disk instabilities and minor mergers may be responsible 
for the bulge growth of targeted galaxies.   

\subsection{Implications in the view of quenching}
\label{subsec:quenching}

\begin{figure}
  \begin{center}
    \epsfig{figure=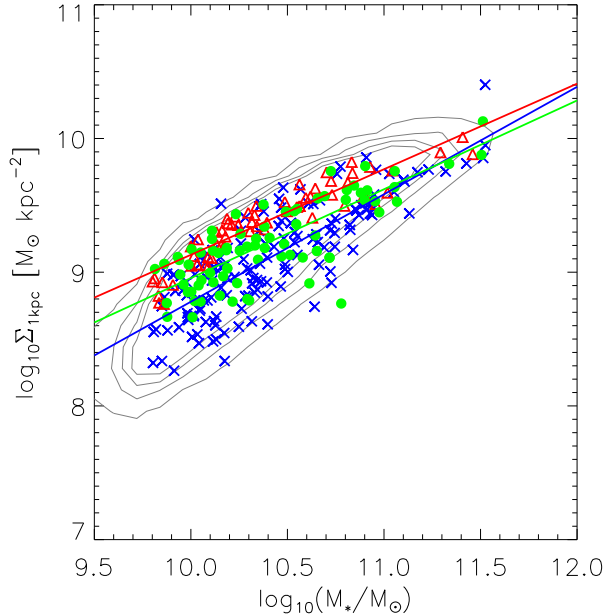,clip=true,width=0.5\textwidth}
  \end{center}
  \caption{$\Sigma_{1kpc}$ vs. stellar mass for BC sample (blue crosses), targeted galaxies (green dots) 
  and RC sample (red triangles). The red solid line is taken from \cite{Fang-13}. The green and blue
  lines are the best fits to the targeted galaxies and BC galaxies. 
  For comparison, a volume-limited sample selected from SDSS is shown in the background contours. }
  \label{fig:quenching}
\end{figure}

The targeted galaxies are found to assemble their central stellar mass rapidly (see Figure \ref{fig:SFH_profile}), 
indicating a rapid (pseudo) bulge growth. 
Massive bulge is believed to be the key factor for quenching star formation for local galaxies
 \citep{Bell-12, Fang-13}. \cite{Bluck-14} found that the bulge mass is more critical
 in quenching star formation rather than bulge to total stellar mass ratio for local galaxies. 
More recently, \cite{Barro-17} investigated the evolution of the stellar surface mass density within central 
1 kpc ($\Sigma_{\rm 1kpc}$) in SF galaxies and quiescent galaxies since redshift of 3. They found 
the growth of dense stellar core is the necessary condition for star formation quenching. 

Figure \ref{fig:quenching} shows $\Sigma_{\rm 1kpc}$ versus stellar mass 
for BC sample (blue crosses), targeted sample (green dots)  and RC sample (red triangles). 
The red solid line is taken from \cite{Fang-13} for quenched galaxies. 
For comparison, $\Sigma_{\rm 1kpc}$ used here is computed following 
the same procedure in \cite{Fang-13}. 
The green and blue lines are the best fits of the targeted galaxies and BC galaxies.  
For comparison, a volume-limited sample selected from SDSS is shown in the 
background contours (the same dataset in Figure \ref{fig:global_properties}).
We find the targeted galaxies are broadly located between two control samples on 
the diagram of $\Sigma_{\rm 1kpc}$-\mstar. 
Considering their intense star formation activities at the center,  
they are undergoing rapid (pseudo) bulge growth, and will meet the quenching requirements 
on a more rapid timescale than normal SF galaxies. 

It is well known that the bulge mass has a tight relation with the central massive black hole mass \citep[e.g.][]{Hu-08,Daddi-07,McConnell-Ma-13}, 
suggesting the coevloution of bulge growth and black hole accretion. The AGN feedback is supposed to be 
one of mechanisms for quenching star formation in galaxies \citep[e.g.][]{Kauffmann-04, Schawinski-07, Dekel-Burkert-14}.  
We find 4 ($\sim$5\%) targeted galaxies are classified 
as Seyfert galaxies according to the BPT diagram \citep{Baldwin-Phillips-Terlevich-81, Kauffmann-03b}, 
while only 3 ($\sim$2\%) BC galaxies are classified as Seyfert galaxies. 
Because of the limitation of sample size, we do not overinterpret this result. 

\section{Summary}
\label{sec:summary}

In this paper, we have for the first time selected a sample of massive SF galaxies with outside-in
stellar mass assembly mode and analyzed their properties and environments. 
We made use of the released data from the largest on-going IFS survey, MaNGA project, which contains 
1390 galaxies. We have performed the spectral fittings by using 
 two methods: a public spectral fitting code STARLIGHT \citep{CidFernandes-05} and a minimized 
chi-squared code from \cite{Li-05}.  We measured the \dindex, \hda\ index and \ewhae\ for individual 
spaxels of each galaxy,  as well as stellar mass maps from STARLIGHT outputs.

We selected the massive SF galaxies to have \mstar$>10^{9.8}$\msun, \nuvr$<$5 and 
\dindexin$<$\dindexout. Mergers and disturbed galaxies are excluded by visually inspection. 
A total of 77 galaxies are selected under the criteria, which accounts for 14\% of 
massive star-forming galaxies. For comparison, we have built two control samples closely matched in
stellar mass with targeted sample. The BC sample is selected to have \dindexin$>$\dindexout\ 
and \nuvr$<$5, and the RC sample is selected to have \nuvr$>$5.
The main results in this work are listed below. 
  
\begin{itemize}
\item The targeted sample has smaller size and higher concentration than BC sample, and has 
larger size and lower concentration than RC sample.  
In agreement with this, we find an increasing trend of smooth-like fraction and bulge fraction
 from the BC sample, the targeted sample to the RC sample as a whole. 

\item The targeted sample also has larger global SFR (0.36 dex higher) and larger global gas-phase 
metallicity (0.12 dex higher) than BC sample.  Further, 
the median profile of gas-phase metallicity of targeted sample is systematically higher 
than that of the BC sample for $\sim$0.07 dex within 1.5\re.  When plotting the individual spaxels on 
the $\Sigma_*$-metallicity plane, the targeted sample appears to be more metal-rich than the BC sample
 at high $\Sigma_*$ end, and become comparable when $\Sigma_*<$10$^8$ M$_{\odot}$ kpc$^{-2}$. 

\item  The targeted sample has a slightly positive gradient of \dindex,  a slightly negative gradient of \hda\
 index and a pronounced negative gradient of \lgewhae,
  while the BC galaxies show pronounced negative gradients of \dindex, and positive gradients of \hda\
  index and \lgewhae. These are consistent with our selected criteria.  

\item   The median $\Sigma_*$ profiles of three samples are almost parallel. The median $\Sigma_*$
 profile of targeted sample is about 0.22 dex higher than the BC sample, 
 and 0.39 dex lower than the RC sample. 

\item We have analyzed the environment of our samples, including projected 2PCCFs and local density 
distributions ($\sim$2 Mpc).
The targeted sample is more clustered than BC sample as a whole, especially at scales of 0.8-2 Mpc (0.3-0.5 dex higher), and appears to be more weakly clustered than RC sample, especially at the scale of
 50-300 kpc (0.3-0.5 dex lower). However, the wide density distribution (see Figure \ref{fig:overdensity}) 
 indicates a variety of environments for targeted galaxies.

\end{itemize}

The progenitors of the targeted galaxies may be normal SF galaxies suffering from gas inflow,  
or quiescent galaxies with cold gas accretion. 
The median $\Sigma_*$ profile of targeted sample is in between that of two control samples, 
 indicating that the cold gas accretion of quiescent galaxies is not the main approach to 
the formation of the targeted galaxies. 
The structure properties of targeted galaxies are also in between of two control
samples, suggesting that massive SF galaxies with outside-in assembly mode are probably 
in the transitional phase from the normal SF galaxies to quiescent galaxies.  
They are experiencing rapid (pseudo) bulge growth and will meet the requirement of quenching 
on a more rapid timescale than normal SF galaxies. 

It is well known that the existence of bar is efficient to cause gas inflow and intense 
central star formation activities \citep{Wang-12, Lin-17}. Only 14 ($\sim$19\%) targeted galaxies host a bar, 
indicating the existence of the bar is not the main driver for the outside-in assembly mode.
We also found 19 ($\sim$25\%) targeted galaxies are found to have one or more companions within 100 kpc,
indicating the interaction with companions is also not the main reason. The majority of targeted galaxies (60\%),
 appeared to have smooth-like morphologies, are not found to have a bar or 
 companions within 100 kpc.
The formation mechanisms of them are not clear. The disk instabilities in secular evolution and minor mergers
 may be another additional channels for bulge growth and morphological 
 transformation \citep{Porter-14b, Brennan-15}.
In the near future, the on-going large IFS surveys will provide larger sample of galaxies with 
outside-in assembly mode, and help people to understand the origins of outside-in assembly mode. \\
  
\acknowledgments


This work is supported by the National Basic Research Program of China (973 Program)(2015CB857004), 
the National Natural Science Foundation of China (NSFC, Nos. 11320101002, 11522324, 11421303, 
and 11433005) and the Fundamental Research Funds for the Central Universities.
EW acknowledges the support from the Youth Innovation Fund by University of 
Science and Technology of China (No. WK2030220019). 
 
Funding for the Sloan Digital Sky Survey IV has been provided by
the Alfred P. Sloan Foundation, the U.S. Department of Energy Office of
Science, and the Participating Institutions. SDSS-IV acknowledges
support and resources from the Center for High-Performance Computing at
the University of Utah. The SDSS web site is www.sdss.org.

SDSS-IV is managed by the Astrophysical Research Consortium for the
Participating Institutions of the SDSS Collaboration including the
Brazilian Participation Group, the Carnegie Institution for Science,
Carnegie Mellon University, the Chilean Participation Group, the French Participation Group, 
Harvard-Smithsonian Center for Astrophysics,
Instituto de Astrof\'isica de Canarias, The Johns Hopkins University,
Kavli Institute for the Physics and Mathematics of the Universe (IPMU) /
University of Tokyo, Lawrence Berkeley National Laboratory,
Leibniz Institut f\"ur Astrophysik Potsdam (AIP),
Max-Planck-Institut f\"ur Astronomie (MPIA Heidelberg),
Max-Planck-Institut f\"ur Astrophysik (MPA Garching),
Max-Planck-Institut f\"ur Extraterrestrische Physik (MPE),
National Astronomical Observatory of China, New Mexico State University,
New York University, University of Notre Dame,
Observat\'ario Nacional / MCTI, The Ohio State University,
Pennsylvania State University, Shanghai Astronomical Observatory,
United Kingdom Participation Group,
Universidad Nacional Aut\'onoma de M\'exico, University of Arizona,
University of Colorado Boulder, University of Oxford, University of Portsmouth,
University of Utah, University of Virginia, University of Washington, University of Wisconsin,
Vanderbilt University, and Yale University.

\bibliography{rewritebib.bib}

\appendix
\section{Stellar age profiles from STARLIGHT measurements}

\begin{figure*}
  \begin{center}
    \epsfig{figure=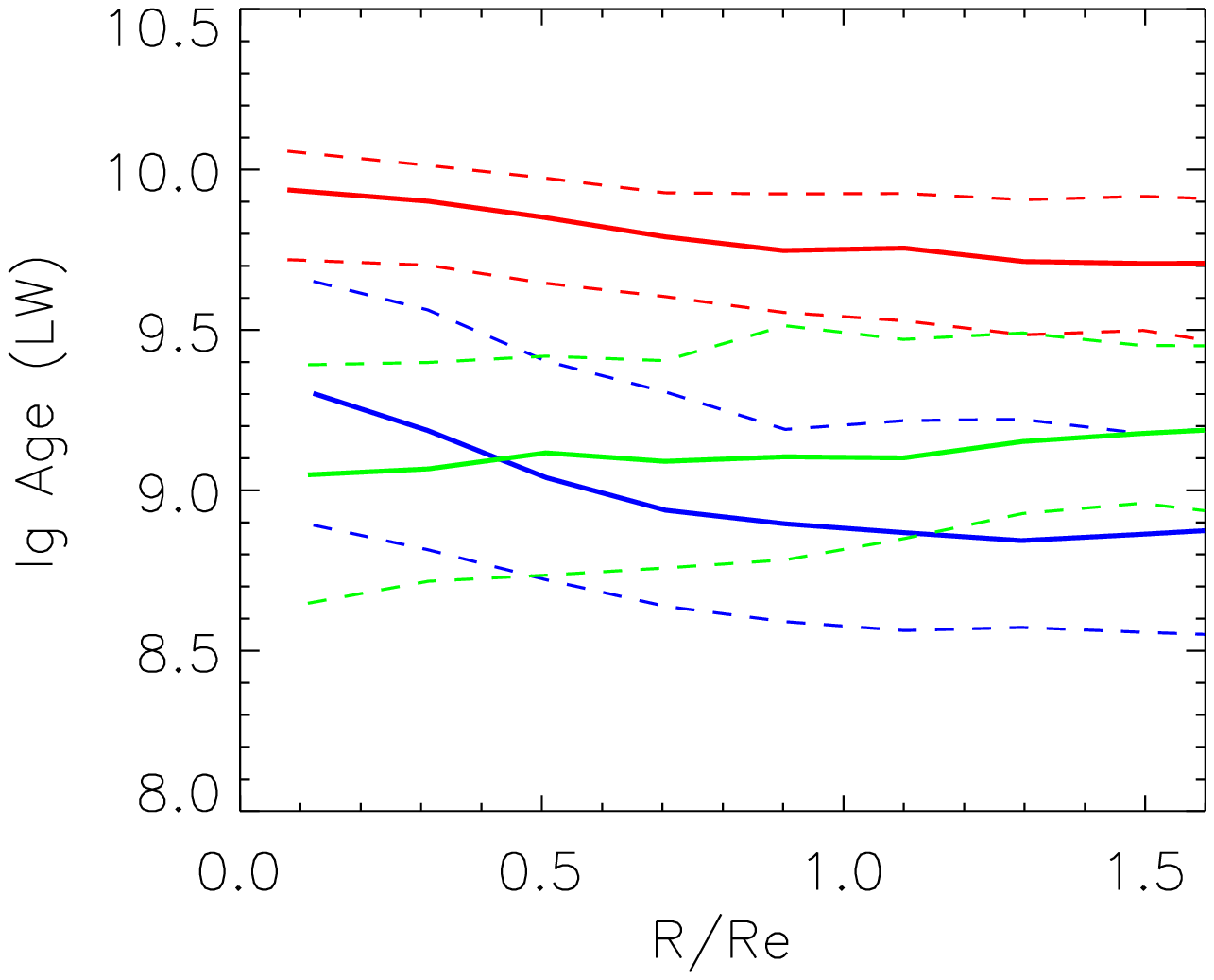,clip=true,width=0.40\textwidth}
    \epsfig{figure=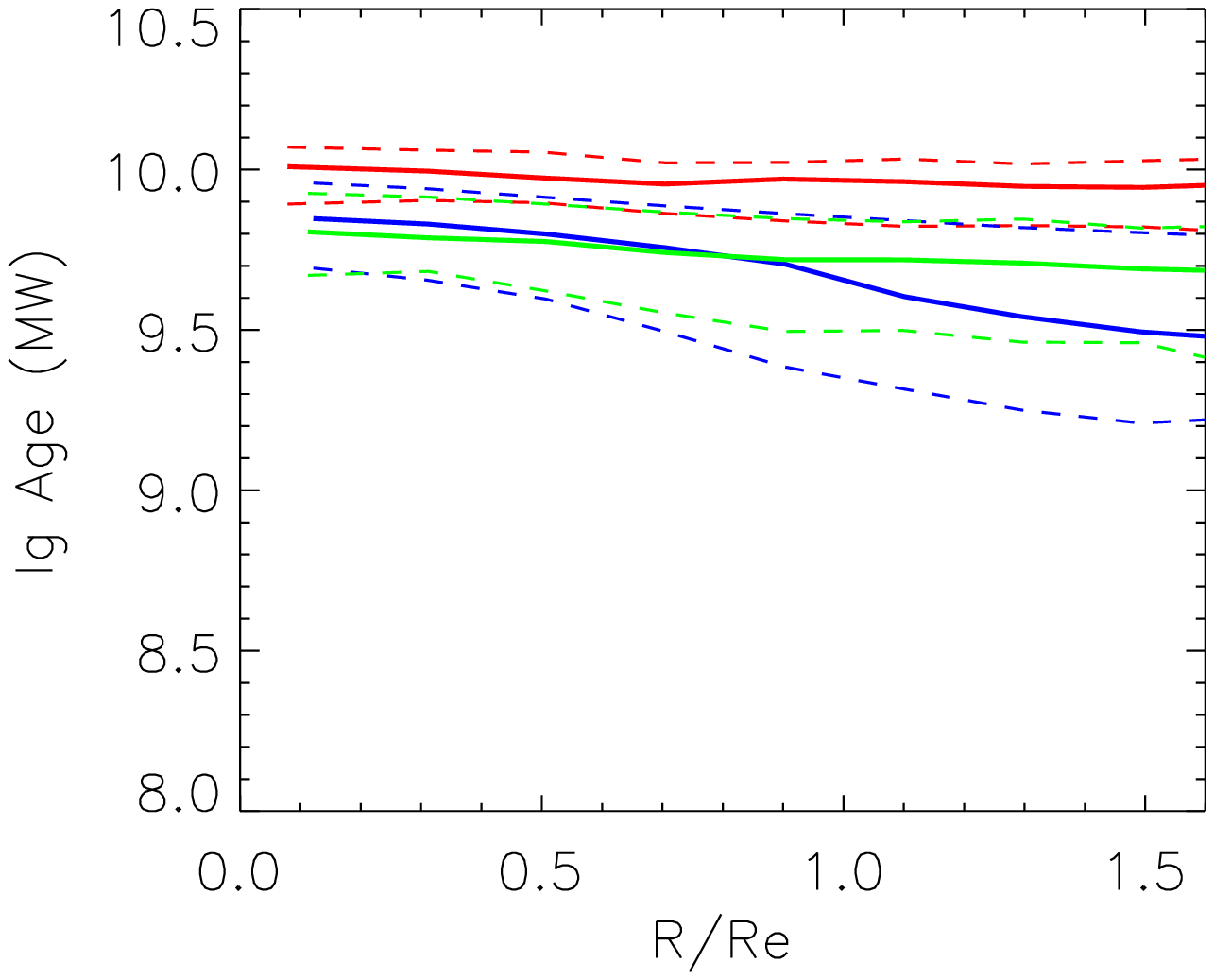,clip=true,width=0.40\textwidth}
  \end{center}
  \caption{The median profiles of light-weighted (left panel) and mass-weighted (right panel) stellar ages for 
  targeted (green curve),  BC (blue curve) and RC (red curve) samples. In each panel, the dashed lines correspond 
  to the 20-80 per cent percentile range.}
  \label{fig:age_profile}
\end{figure*}

Since our main results are based on the three diagnostic parameters, we have checked our results 
by using the mean stellar age from STARLIGHT outputs. The left panel of Figure \ref{fig:age_profile} shows 
median profiles of the light-weighted stellar age 
for targeted (green curve), BC (blue curve) and RC (red curve) samples, 
while the right panel shows median profiles of the mass-weighted stellar age for these three samples. 
In each panel, the dashed lines correspond to the 20-80 per cent percentile range.
The mass-weighted ages trace the star formation burst at early stages, while the light-weighted 
ages trace the recent star formation burst. 
As shown, the targeted sample exhibits a slightly positive gradient of light-weighted age, and a slightly negative
gradient of mass-weighted age. This supports our conclusion that the targeted galaxies suffer a reactivation of 
their central star formation at recent times.

\label{lastpage}
\end{document}